\documentclass[10pt,a4paper]{article}
\usepackage{epsfig}
\begin{document}
\textwidth=135mm
 \textheight=200mm
\def\Journal#1#2#3#4{//{#1}\ #4.\ {V.#2.}\ P.#3. }
\def\AP{{  Ann. Phys.}}
\def\AA{{  Astron. Astrophys.}}
\def\AAT{{  Astron. Astrophys. Trans.}}
\def\AFZ{{  Astrofizika}}
\def\AJ{{  Astron. J.}}
\def\APJ{{  Astrophys. J.}}
\def\APJS{{  Astrophys. J. Suppl.}}
\def\MATH{{  J. Math. Phys.}}
\def\MNRAS{{  Month. Not. Roy. Astron. Soc.}}
\def\N{{  Nature}}
\def\NA{{  New Astron.}}
\def\NPA{{  Nucl. Phys.} A.}
\def\NPB{{  Nucl. Phys.} B.}
\def\NCA{{  Nuovo Cimento} A.}
\def\PHYS{{  Physica}}
\def\PLA{{  Phys. Lett.} A.}
\def\PLB{{  Phys. Lett.} B.}
\def\PLD{{  Phys. Lett.} D.}
\def\PL{{  Phys. Lett.}}
\def\PRL{  Phys. Rev. Lett.}
\def\PREV{  Phys. Rev.}
\def\PREP{  Phys. Rep.}
\def\PRA{{  Phys. Rev.} A.}
\def\PRD{{  Phys. Rev.} D.}
\def\PRC{{  Phys. Rev.} C.}
\def\PRB{{  Phys. Rev.} B.}
\def\PRO{{  Prog. Theor. Phys.}}
\def\SC{{  Science}}
\def\ZPC{{  Z. Phys.} C.}
\def\ZPA{{  Z. Phys.} A.}
\def\ANNP{  Ann. Phys. (N.Y.)}
\def\RMP{{  Rev. Mod. Phys.}}
\def\CHEM{{  J. Chem. Phys.}}
\def\INT{{  Int. J. Mod. Phys.} E.}
\def\arx{{  arXiv:}}
\def\r{\vec r}
\def\R{\vec R}
\def\p{\vec p}
\def\P{\vec P}
\def\q{\vec q}
\def\ss{\mbox{\boldmath $\sigma$}}
\newcommand{\be}{\begin{equation}}
\newcommand{\ee}{\end{equation}}
\newcommand{\bea}{\begin{eqnarray}}
\newcommand{\eea}{\end{eqnarray}}
\newcommand{\nn}{\nonumber}
\newcommand{\gsim}{\stackrel{\scriptstyle >}{\phantom{}_{\sim}}}
\newcommand{\lsim}{\stackrel{\scriptstyle <}{\phantom{}_{\sim}}}
\newcommand\eqn[1]{(\ref{#1})}
\newcommand\eqns[2]{(\ref{#1})-(\ref{#2})}
\newcommand{\pslash}{{p\!\!\!/}}
\newcommand{\dslash}{{\partial\!\!\!/}}
\newcommand{\ie}{, {\it i.e.}, }
\newcommand{\eg}{, {\it e.g.}, }
\newcommand{\figurewidth}{\linewidth}

\begin{center}
{\bfseries Structure and cooling of compact stars
\footnote{\small Lecture notes for the Helmholtz International Summer School
on "Dense Matter in Heavy-Ion Collisions and Astrophysics", JINR, Dubna,
August 21 - September 1, 2006.}}

\vskip 5mm

H. Grigorian$^{\dag,\ddag}$

\vskip 5mm

{\small {\it $^\dag$ Laboratory for Information Technologies at JINR,
141980 Dubna, Russia}} \\
{\small {\it $^\ddag$ Department of Physics, Yerevan State
University, 375025 Yerevan, Armenia}}
\\

\end{center}

\vskip 5mm

\centerline{\bf Abstract}
We study the structure and evolution of neutron stars (NS) the interiors
of which are modeled using microscopic approaches and constrained by the
condition that the equation of state (EoS) of matter extrapolated to high
densities should not contradict known observational data from compact stars
and experimental data from heavy-ion collisions (HIC).
We use modern cooling simulations to extract distributions of NS masses
required to reproduce those of the yet sparse data in the Temperature-Age
(TA) plane. By comparing the results with a mass distribution for
young, nearby NSs used in population synthesis we can sharpen the
NS cooling constraints.

\vskip 5mm

\section{\label{sec:intro}Introduction}

Among compact stars denoted here generically as NSs one can
theoretically distinguish three main classes according to their
composition: hadronic stars, quark stars (absolutely bare or with
tiny crusts), and hybrid stars (HyS).
This classification is based on the understanding of the EoS of matter under
extreme conditions which is at present lively debated.
Can astronomy of NSs together with experimental data from HICs, where
the matter exceeds the nuclear saturation density allow us to
identify the true "name" of a given observed compact star? To
answer this "simple" question one needs to do a very complex
job. We have to collect experimentally justified models of nuclear
matter, extrapolate their properties to sufficiently high
densities, make a compact object model based on that and return to
observations for comparison of the results. Probably this cycle is
only one loop of iteration like in any perturbation approach.

To be sure that one moves in the correct direction one needs to
define constraints on our modeling from the investigation of
high-density behavior of nuclear matter (NM) activated in such an
experiments as planned when the new accelerator facility (FAIR) at
GSI Darmstadt will be constructed. On the other hand the new
observations in astronomy provide new limits for the mass and the
mass-radius relationship, surface temperature etc. of compact
stars. These data constitute stringent constraints on the equation
of state of strongly interacting matter at high densities, see
\cite{Klahn:2006ir} and references therein.
Of course experiments such as CBM (=compressed baryon matter) at FAIR
or NICA at JINR which are dedicated for the investigation of the phase
transition from hadronic matter to the quark-gluon plasma (QGP) in HIC
have a natural interest for the theory of NSs and particularly for the
question whether quark matter exists in the core of a NS.

Here in these lectures, we demonstrate that the present-day
knowledge of hydrodynamical properties of dense matter allows to
construct hybrid EsoS with a critical density of the deconfinement
phase transition low enough to allow for extended quark matter
cores and stiff enough to comply with the new mass measurements of
compact stars \cite{Klahn:2006iw}.
As has been argued by Alford et al. \cite{Alford:2006vz}
\cite{Alford:2004pf} there are several modern QCD-motivated quark
matter EsoS which could provide enough stiffness of high-density
matter to be not in conflict with the new mass and mass-radius
constraints.
When compared with hadronic EsoS it turns out that quark and hadron matter
have rather similar hydrodynamical properties in the region of the
deconfinement transition.
This means that hybrid stars can masquerade as neutron
stars once the parameters of a generic phenomenological quark
matter EoS have been chosen appropriately \cite{Alford:2004pf,Klahn:2006iw}.

However, we suggest a new tool for ``unmasking'' the composition
of neutron star interiors which is based on the fact that the
state of matter at high densities determines the statistics of
both NS observables, the temperature-age (TA) data as well as the
mass distribution.

\section{Equation of state of neutron star matter}

\subsection{Hadronic EoS}

There are numerous comparative studies of NM approaches for HIC
and NS physics applications in which a representation of the NM
EoS has been employed which is based on the nucleonic part of the
binding energy per particle given in the form
\begin{equation}
\label{eq:BA} E(n,\beta)=E_0(n) + \beta^2 E_S(n),
\end{equation}
where $\beta = 1 - 2 x$ is the asymmetry parameter depending on
the proton fraction $x=n_p/n$ with the total baryon density
$n=n_n+n_p$. In Eq.~(\ref{eq:BA}) the function $E_0(n)$ is the
binding energy in symmetric NM (SNM), and $E_S(n)$ is the
(a)symmetry energy, i.e. the energy difference between pure
neutron matter and SNM. Both contributions $E_0(n)$ and $E_S(n)$
are easily extracted from a given EoS for the cases $\beta=0$ and
$\beta=1$, respectively. The parabolic interpolation has been
widely used in the literature, see e.g. \cite{Lattimer:2000nx}, and appears
rather robust.
From Eq.~(\ref{eq:BA}) all zero temperature EsoS of nonstrange NM can be
derived by applying simple thermodynamic identities \cite{BaBu00}.
In particular, we obtain
\begin{eqnarray}
\varepsilon_B(n,\beta)&=&nE(n,\beta), \\
P_B(n,\beta)&=&n^2\frac{\partial}{\partial n} E(n,\beta),
\end{eqnarray}
\vspace{-0.7cm}
\begin{eqnarray}
\label{eq:munp} \mu_{n,p}(n,\beta)&=& \left(
1+n\frac{\partial}{\partial n} \right) E_0(n) - \left(
\beta^2\mp2\beta-\beta^2n\frac{\partial}{\partial n} \right)
E_S(n)
\end{eqnarray}
for the baryonic energy density $\varepsilon(n)$ and pressure
$P(n)$ as well as the chemical potentials of neutron $\mu_{n}$
(upper sign) and proton $\mu_{p}$ (lower sign), respectively.

The neutron star matter (NSM) has to fulfill the two essential
conditions of $\beta$-equilibrium
\begin{eqnarray}
\label{eq:betaeq} \mu_n = \mu_p + \mu_e = \mu_p + \mu_\mu~~,
\end{eqnarray}
and charge neutrality
\begin{eqnarray}
\label{eq:neutr} n_p -n_e-n_\mu =0~,
\end{eqnarray}
where $\mu_e$ and $\mu_\mu$ are the electron and muon chemical
potentials, conjugate to the corresponding densities $n_e$ and
$n_\mu$.  In case of a first order phase transition a
mixed phase could arise in some density interval, see
\cite{Glendenning:1992vb}. In general, the local charge neutrality
condition could be replaced by the global one. However, due to the
charge screening for the phase transition to quark matter this
density interval is essentially narrowed
\cite{Voskresensky:2002hu,Maruyama:2005tb}. The effect of the
mixed phase on the EoS is also minor.

Due to Eq.~(\ref{eq:betaeq}) the  chemical potentials for muons
and electrons are equal, $\mu_\mu = \mu_e$. The threshold of muons
to appear in the system is determined by their mass.
In NSM the baryonic and the leptonic parts are considered as an ideal
mixture in $\beta$-equilibrium
\begin{eqnarray}
\label{eq:nsm} \varepsilon(n, \beta)&=&\varepsilon_B(n, \beta)
+\varepsilon_e(n, \beta)+\varepsilon_\mu(n, \beta)~,
\\
P(n,\beta)&=&P_B(n,\beta)+P_e(n,\beta)+P_\mu(n,\beta)~.
\end{eqnarray}

Under NS conditions one parameter such as the baryo-chemical potential
$\mu_b$ or the baryon density conjugate to it is sufficient for a
complete description. In $\beta$-equilibrated neutron matter the chemical
potential of the neutrons equals the baronic one, $\mu_n = \mu_b$.
Applying Eq.~(\ref{eq:munp}) and
Eq.~(\ref{eq:betaeq}) shows that the electron and muon chemical
potential can be written as an explicit function of baryon density
and asymmetry parameter,
\begin{equation}
\label{eq:mue} \mu_e(n,\beta)=4\beta E_S(n).
\end{equation}
Both electrons and muons are described as a massive, relativistic
ideal Fermi gas.

With the above relations only one degree of freedom, namely the
baryon density, remains in charge neutral and $\beta$-equilibrated
NSM at zero temperature. Within this description actual properties
of NM depend on the behavior of $E_0(n)$ and $E_S(n)$ only. Both
can be deduced easily from any EoS introduced in these lectures.

Here in the discussions we use the nuclear EsoS that originate
from relativistic descriptions of NM, because in the range of
densities relevant NSs and HICs relativistic effects are
important.

A brief description of different approaches is given in the following.
For more details, see\cite{Klahn:2006ir} and references therein.

\begin{itemize}
  \item {\it Phenomenological models}\\
originate from \cite{Wal74} and are based on a relativistic mean-field
(RMF) description of NM with nucleons and mesons as degrees of
freedom.

The mesons couple minimally to the nucleons. The coupling
strengths are adjusted to properties of NM or atomic nuclei. A
scalar meson ($\sigma$) and a vector meson ($\omega$)  are treated
as classical fields generating scalar and vector interactions. The
isovector contribution is generally represented by a vector meson
$\rho$. In order to improve the description of experimental data,
a medium dependence of  the effective interaction has to be
incorporated into the model.

In many applications of the RMF model, non-linear (NL)
self-interactions of the $\sigma$ meson were introduced with
considerable success and were later extended to other meson fields
(see \cite{Tok98}).  As an alternative to NL RMF models, appoaches with
density-dependent nucleon-meson couplings were developed (see
\cite{Lal05} and references within), where for a flexible
description of the medium dependence several parameterizations
were introduced. In the NL$\rho$ the isovector part of the
interaction is described, as usual, only by a $\rho$ meson. The
set NL$\rho\delta$ also includes a scalar isovector meson
$\delta$.

 The density
dependent RMF models are also represented here by
\cite{Typel:2005ba} (DD models). They are obtained from a fit to
properties of finite nuclei (binding energies, radii, surface
thickenesses, neutron skins and spin-orbit splittings).
 In the D${}^{3}$C model additional
couplings of the isoscalar mesons to derivatives of the nucleon
field are introduced.

\item {\it Dirac-Brueckner-Hartree-Fock (DBHF) approach}\\

One of the microscopic approaches starting {\it ab initio} from a
given free nucleon-nucleon interaction that is fitted to
nucleon-nucleon scattering data and deuteron properties is the DBHF
method \cite{honnef}.
In this approach the nucleon  inside the medium is dressed by the
self-energy $\Sigma$ based on a T-matrix. The in-medium T-matrix
which is obtained from the Bethe-Salpeter equation plays the role
of an effective two-body interaction which contains all
short-range and many-body correlations in the ladder
approximation.  It is possible to extract the nucleon
self-energies from DBHF calculations which can be compared with
the corresponding quantities in phenomenological RMF models, but
this is not completely unambiguous as discussed in
Ref.~\cite{DaFuFae05}. Here, we use recent results of (asymmetric)
NM calculations in the DBHF approach with the relativistic Bonn A
potential in the subtracted T-matrix representation.

  \item {\it A bridging approach between fully microscopic and more
phenomenological descriptions}\\
 that adjusts the parameters of
the RMF model to results extracted from microscopic approaches. As
an example of this method, we use a nonlinear RMF model (KVR) (or
slightly modified parameter set (KVOR)) with couplings and meson
masses depending on the $\sigma$- meson field \cite{KoVo05}. The
parameters were adjusted to describe the SNM and NSM EoS of the
Urbana-Argonne group \cite{AkPaRa98}  at densities below four
times the saturation density. In these models not only the nucleon
but also the meson masses decrease with increasing NM density.
Being motivated by the Brown-Rho scaling assumption, see
\cite{Brown:1991kk}, and the equivalence theorem between different
RMF schemes, these models use only one extra parameter compared to
the standard NL RMF model (NL model).

\end{itemize}

The sets of parameters and references of used models are given in
\cite{Klahn:2006ir}.

\begin{figure}[htb]
\includegraphics[width=7cm,height=12cm, angle=-90]{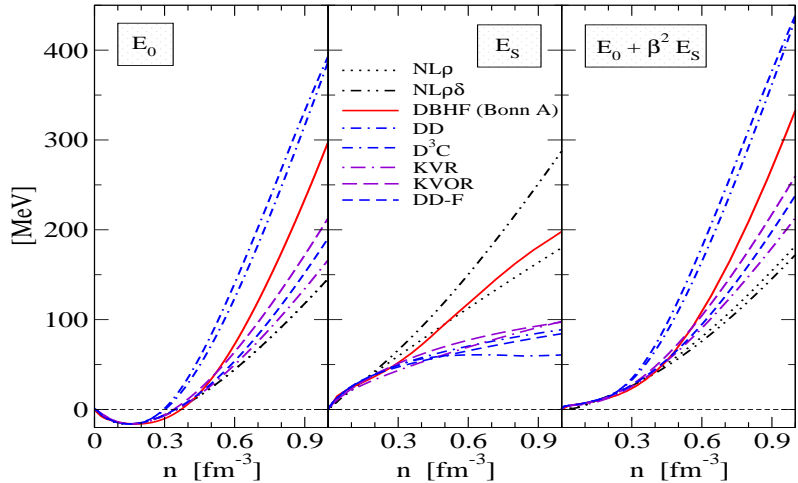}
\caption{\label{fig:e0_s0_nsm} The energy per nucleon in SNM
$E_{0}(n)$ (left panel), the symmetry energy $E_{S}(n)$ (middle
panel) and the energy per nucleon in NSM ($\beta$-equilibrated and
charge neutral) for the investigated models (right panel). }
\end{figure}

The variation in the NM parameters is directly reflected in the
behavior of the energy per nucleon in SNM $E_{0}(n)$ and of the
symmetry energy $E_{S}(n)$ at densities above saturation as shown
in Fig.~\ref{fig:e0_s0_nsm}. The various models of this study
predict considerably different values for $E_{0}(n)$ and
$E_{S}(n)$ at high densities. Under the condition of
$\beta$-equilibrium, however, the range of binding energy per
nucleon $E(n,\beta)$ shows a much smaller variation than expected
from $E_{0}(n)$ and  $E_{S}(n)$. This is shown in the right panel
of Fig.~\ref{fig:e0_s0_nsm}.

\subsection{Quark matter EoS}

Since there are no solutions of QCD for quark matter in the
nonperturbative domain close to the chiral/ deconfinement
transition at zero temperature and finite density one needs an
effective model for the quark matter EoS satisfying at least the
symmetry requirements of QCD.
In order to discuss whether deconfined quarks
can exist in neutron stars a successful class of approaches is that
of Nambu--Jona-Lasinio (NJL) type models having a Lagrangian
with chiral symmetry which is dynamically broken in the
nonperturbative vacuum. In our applications in astrophysics we use
this microscopic approach, see\cite{Buballa:2003qv} and references
therein.

In present investigation we took a three-flavor chiral quark model
with self-consistently determined quark masses and pairing gaps
\cite{Blaschke:2005uj} similar to the parallel developments in
Refs. \cite{Ruster:2005jc,Abuki:2005ms}, but generalizing
\cite{Blaschke:2005uj} by including an isoscalar vector meson
current \cite{Klahn:2006iw} similar to the Walecka model for nuclear matter.
While a stronger scalar diquark coupling leads to a lowering of the
phase transition density, an increase in the vector coupling makes the quark
matter EoS stiffer.

The thermodynamics of the deconfined quark matter phase as
described within a three-flavor quark model of Nambu--Jona-Lasinio
(NJL) type can be investigated systematically by applying techniques of
finite-temperature field theory \cite{Kapusta:2006}.
The path-integral representation of the partition function is given by
\begin{eqnarray}
\label{Z} Z(T,\hat{\mu})&=&\int {\mathcal D}\bar{q}{\mathcal D}q
\exp \left\{\int_0^\beta d\tau\int d^3x\,\left[
        \bar{q}\left(i\dslash-\hat{m}+\hat{\mu}\gamma^0\right)q+
{\mathcal L}_{\rm int}
\right]\right\},
\label{Zqqbar} \\
{\mathcal L}_{\rm int} &=& G_S\Bigg[
        \sum_{a=0,3,8}(\bar{q}\tau_aq)^2 -
        \eta_V(\bar{q}\gamma^0q)^2 \nonumber \\
        &&+\eta_D\!\!\!\!\sum_{A=2,5,7}\!\!\!
        (\bar{q}i\gamma_5\tau_A\lambda_AC\bar{q}^T)
        (q^TiC\gamma_5\tau_A\lambda_Aq),
\Bigg],
\end{eqnarray}
where $\hat{\mu}$ and $\hat{m}={\rm diag}_f(m_u,m_d,m_s)$ are the
diagonal chemical potential and current quark mass matrices. For
$a=0$, $\tau_0=(2/3)^{1/2}{\mathbf 1}_f$, otherwise $\tau_a$ and
$\lambda_a$  are Gell-Mann matrices acting in flavor and color
spaces, respectively. $C=i\gamma^2\gamma^0$ is the charge
conjugation operator and $\bar{q}=q^\dagger\gamma^0$. $G_S$,
$\eta_V$, and $\eta_D$ determine the coupling strengths of the
interactions.

The Lorentz three-current, $\bar{q}\vec{\gamma} q $, vanishes
in the static ground state of matter and is therefore omitted
in the vector interaction term in (\ref{Zqqbar}), for further
details see \cite{Klahn:2006iw,Blaschke:2005uj}.

After bosonization using Hubbard-Stratonovich transformations, we
obtain an exact transformation of the original partition function
\eqn{Zqqbar}. In mean-field approximation, when the bosonic
functional integrals are omitted and the collective fields are
fixed at the extremum of the action we obtain
\begin{eqnarray}
\Omega_{MF}(T,\mu) &=&
        -\frac{1}{\beta V}\ln Z_{MF}(T,\mu)\nonumber\\
&=& \frac{1}{8 G_S}\left[\sum_{i=u,d,s}(m^*_i-m_i)^2
  - \frac{2}{\eta_V}(2\omega_0^2+\phi_0^2)
        +\frac{2}{\eta_D}\sum_{A=2,5,7}|\Delta_{AA}|^2\right]
        \nonumber \\
&&-\int\frac{d^3p}{(2\pi)^3}\sum_{a=1}^{18}
        \left[E_a+2T\ln\left(1+e^{-E_a/T}\right)\right]
        + \Omega_l - \Omega_0~.
\label{potential}
\end{eqnarray}
Here, $\Omega_l$ is the thermodynamic potential for electrons and
muons, and $\Omega_0$ is a divergent term that is subtracted in
order to get zero pressure and energy density in vacuum
($T=\mu=0$). $E_a(p)$ are the quasiparticle dispersion relations.
Here, $\Delta_{AA}$ are the diquark gaps. $\hat{m}^*$ is the
diagonal renormalized mass matrix and $\hat{\mu}^*$ the
renormalized chemical potential matrix, $\hat{\mu}^*={\rm diag}_f
(\mu_u-G_S\eta_V\omega_0,\mu_d-G_S\eta_V\omega_0,\mu_s-G_S\eta_V\phi_0)$.
The gaps and the renormalized masses are determined by
minimization of the mean-field thermodynamic potential
\eqn{potential}, subject to charge neutrality constraints which
depend on the application we consider. In the (approximately)
isospin symmetric situation of a heavy-ion collision, the color
charges are neutralized, while the electric charge in general is
non-zero. For matter in $\beta$-equilibrium, also the electric
charge is neutralized. Further details are given in Refs.
\cite{Klahn:2006iw,Blaschke:2005uj,Ruster:2005jc,Abuki:2005ms}.

\begin{figure}
\centerline{\psfig{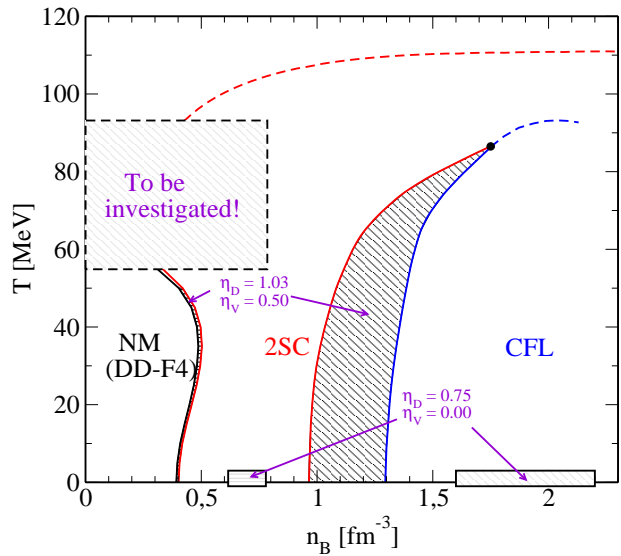}
} \caption{Phase diagram for isospin symmetry using the most
favorable hybrid EoS of the present study. The NM-2SC phase
transition is almost a crossover. The model DD-F4 is used as a
finite-temperature extension of DBHF. For the parameter set
($\eta_D=0.75$, $\eta_V=0.0$) the flow constraint is fulfilled but
no stable hybrid stars are obtained.
\label{image6}}
\end{figure}

We consider $\eta_D$ as a free parameter of the quark matter
model, to be tuned with the present phenomenological constraints
on the high-density EoS. Similarly, the relation between the
coupling in the scalar and vector meson channels, $\eta_V$, is
considered as a free parameter of the model. The remaining degrees
of freedom are fixed according to the NJL model parameterization
in Table I of \cite{Grigorian:2006qe}, where a fit to low-energy
phenomenological results has been made.

 For our investigations of hybrid star
structure we use two hadronic matter EsoS, where DBHF is an
ab-initio calculation for the Bonn-A nucleon-nucleon potential
within the Dirac-Brueckner-Hartree-Fock approach \cite{DaFuFae05},
discussed in the context of compact star constraints in
\cite{Klahn:2006ir,Grigorian:2006pu}. DD-F4 denotes a relativistic
mean-field model of the EoS with density-dependent masses and
coupling constants adjusted to mimick the behavior of the DBHF
approach \cite{Typel:2005ba}. The transition to the quark matter
phase is obtained by a Maxwell construction, where the critical
chemical potential of the phase transition is obtained from the
equality of hadronic and quark matter pressures. A discussion of
the reliability of the Maxwell construction for the case of a set
of conserved charges is given in \cite{Voskresensky:2002hu}.
The resulting phase diagram for the caseof symmetric NM,
relevant for applications in HIC experiments as CBM @ FAIR,
is shown in Fig. \ref{image6}.

\subsection{Constraints on symmetric matter from HIC}

The flow data analysis of dense SNM probed in HICs \cite{DaLaLy02}
reveals a correlation to the stiffness of the EoS which can be
formulated as a constraint to be fulfilled within the testing
scheme introduced here.

\begin{figure}[thb]
\centerline{
\includegraphics[width=7cm, height=7cm,angle=-90]{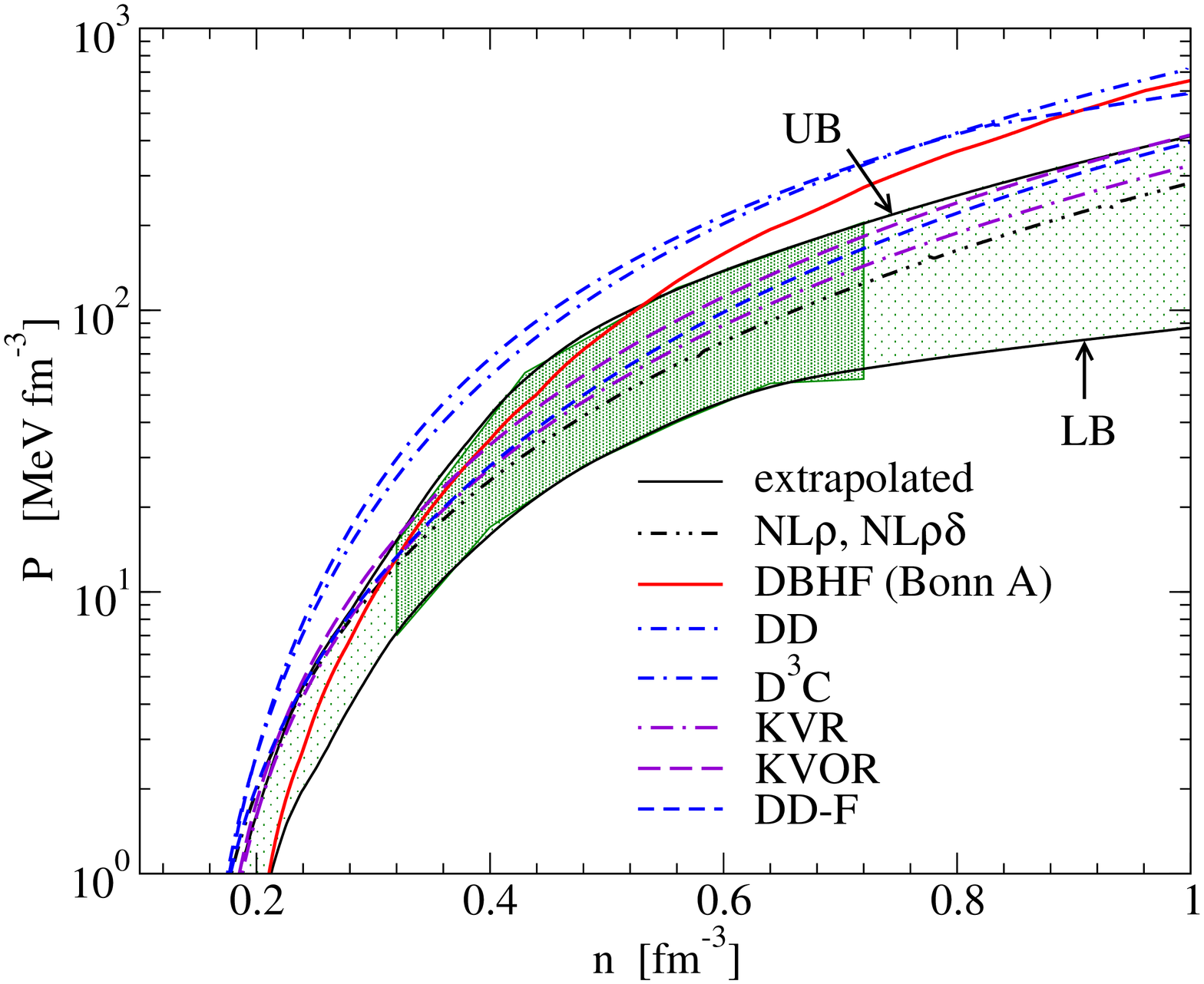}
\hspace{-1cm}\includegraphics[width=7cm,
height=7cm,angle=-90]{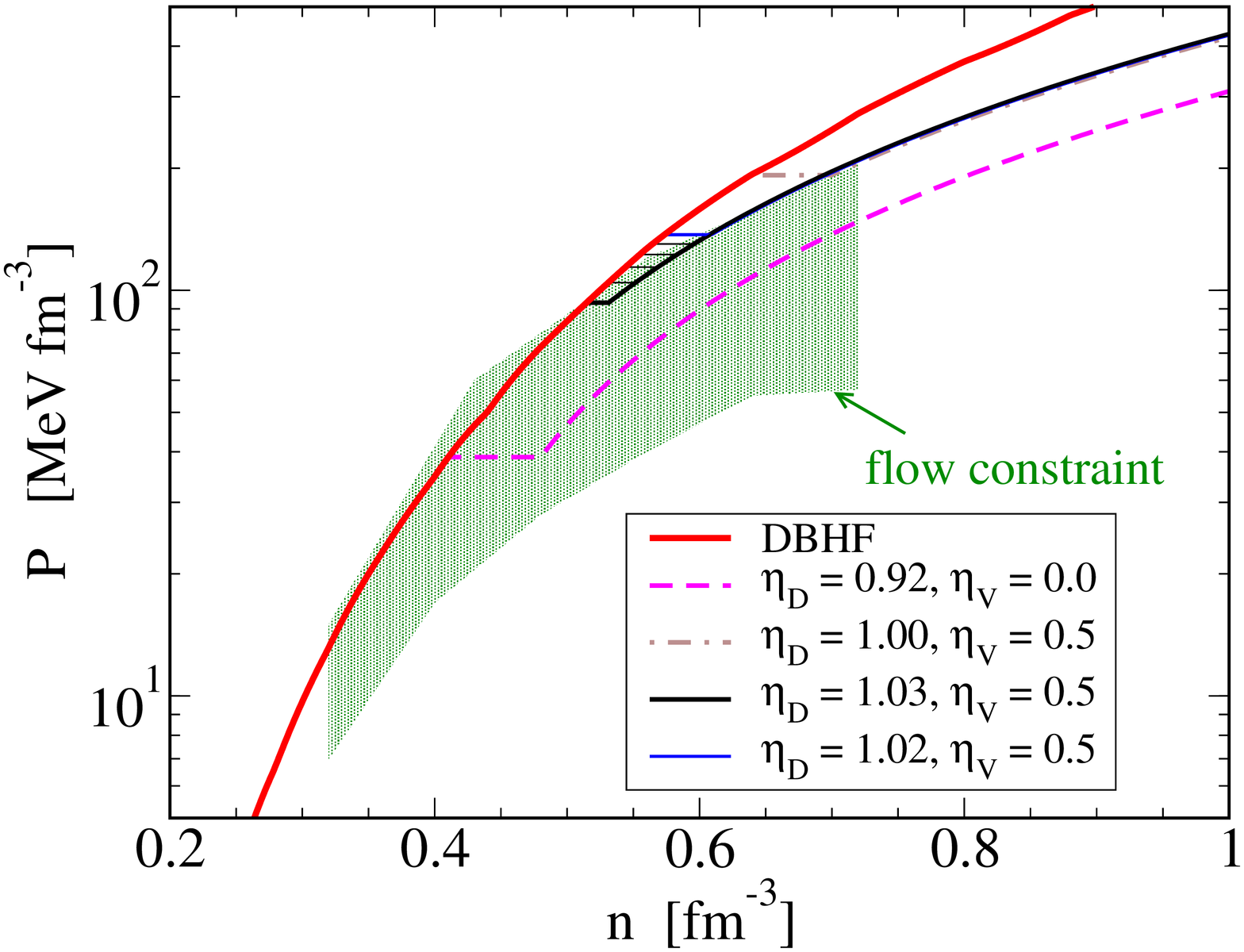}} \caption{
\label{fig:flow_extrapol} Pressure vs. density region consistent with
experimental flow data in SNM (dark shaded region). The light
shaded region on the left panel extrapolates this region to higher
densities within an upper (UB) and lower border (LB).
A set of hadronic EoS is shown for comparison in the left panel, see
text and Ref. \cite{Klahn:2006ir}.
The right panel shows a comparison to EsoS with a deconfinement phase
transition from DBHF hadronic matter to superconducting quark matter with
isovector mean field taken from Ref. \cite{Klahn:2006iw}.
The quark matter EoS favored by the flow
constraint has a vector coupling $\eta_V=0.50$ and a diquark
coupling between $\eta_D=1.02$ (blue solid line) and $\eta_D=1.03$
(black fat solid line); results for four intermediate values
$\eta_D=1.022~\dots~1.028$ are also shown (thin solid lines).}
\label{fig:dan}
\end{figure}

The flow of matter in HICs is directed both forward and
perpendicular (transverse) to the beam axis. At high densities
spectator nucleons may shield the transversal flow into their
direction and generate an inhomogeneous density and thus a
pressure profile in the transversal plane. This effect is commonly
referred to as elliptic flow and depends directly on the given
EoS. An analysis of these nucleon flow data, which depends
essentially only on the isospin independent part of the EoS, was
carried out in a generic model in Ref.~\cite{DaLaLy02}. In
particular it was determined for which range of parameters of the
EoS the model is still compatible with the flow data. The region
thus determined is shown in  Fig.~{\ref{fig:flow_extrapol}} as the
dark shaded region. Ref.~\cite{DaLaLy02} then asserts that this
region limits the range of accessible pressure values at a given
density. For our purposes we extrapolated this region by an upper
(UB) and lower (LB) boundary, enclosing the light shaded region in
Fig.~{\ref{fig:flow_extrapol}}.

 Its upper boundary is expected to be stable against temperature
 variations. The important fact is that the flow constraint
probes essentially only the symmetric part of the binding energy
function $E_0(n)$.

In the hybrid star two phases are separated with a jump in baryon
density corresponding to derivative of the pressure with respect
to the baryochemical potential. The same is shown also for
isospin-symmetric matter in Fig.~\ref{fig:dan}.
A slight variation of the quark matter model parameters  $\eta_D$ and
$\eta_V$ results in considerable changes of the critical density for the
phase transition. The appearent problem of a proper choice of these
parameters we solve by applying the flow constraint shown in
Fig.\ref{fig:dan}.
At first we fix the vector coupling by
demanding that the high density behavior of the hybrid EoS should
be as stiff as possible but still in accordance with the flow
constraint. We obtain $\eta_V=0.50$  such that the problem of the
violation of the flow constraint for the DBHF EoS at high
densities is resolved by the phase transition to quark matter. The
optimal choice for $\eta_D$ is thus between 1.02 and 1.03. In the
following we will investigate the compatibility of the now defined
hybrid star equation of state with CS constraints.

\section{Compact star structure, Mass-Radius relation}

The mass and structure of spherical, nonrotating stars, to which
we limit ourselves in this paper, is calculated by solving the
Tolman-Oppenheimer-Volkov(TOV)-equation, which reads as
\begin{eqnarray}
\label{eq:TOV1} \frac{{\rm d}P(r)}{{\rm d}r} &=&
-~\frac{G[\varepsilon(r)+ P(r)] [m(r)+ 4\pi r^3P(r)]}{r[r-2Gm(r)]} ~,
\end{eqnarray}
where the gravitational mass $m(r)$ inside a sphere of radius $r$
is given by
\begin{eqnarray}
\label{eq:TOV2} m(r)&=& 4\pi\int\limits_0^r{\rm
d}r^\prime\;{r^\prime}^2\varepsilon(r^\prime) ~,
\end{eqnarray}
which includes the effects of the gravitational binding energy.
Eq.~(\ref{eq:TOV1}) describes the gradient of the pressure $P$ and
implicitly the radial distribution of the energy density
$\varepsilon$ inside the star. The EoS, i.e., the relation between
$P$ and $\varepsilon$ leaves as much unknowns as we need to solve
the set of differential equations. We supplement our EsoS
describing the NSs interior by an EoS for the crust. For that we
use a simple BPS model \cite{BaPe71}. Due to uncertainties with
different crust models one may obtain slightly different
mass-radius relations.

\begin{figure}[htb]
\centerline{
\includegraphics[width=7cm,height=7cm, angle=-90]{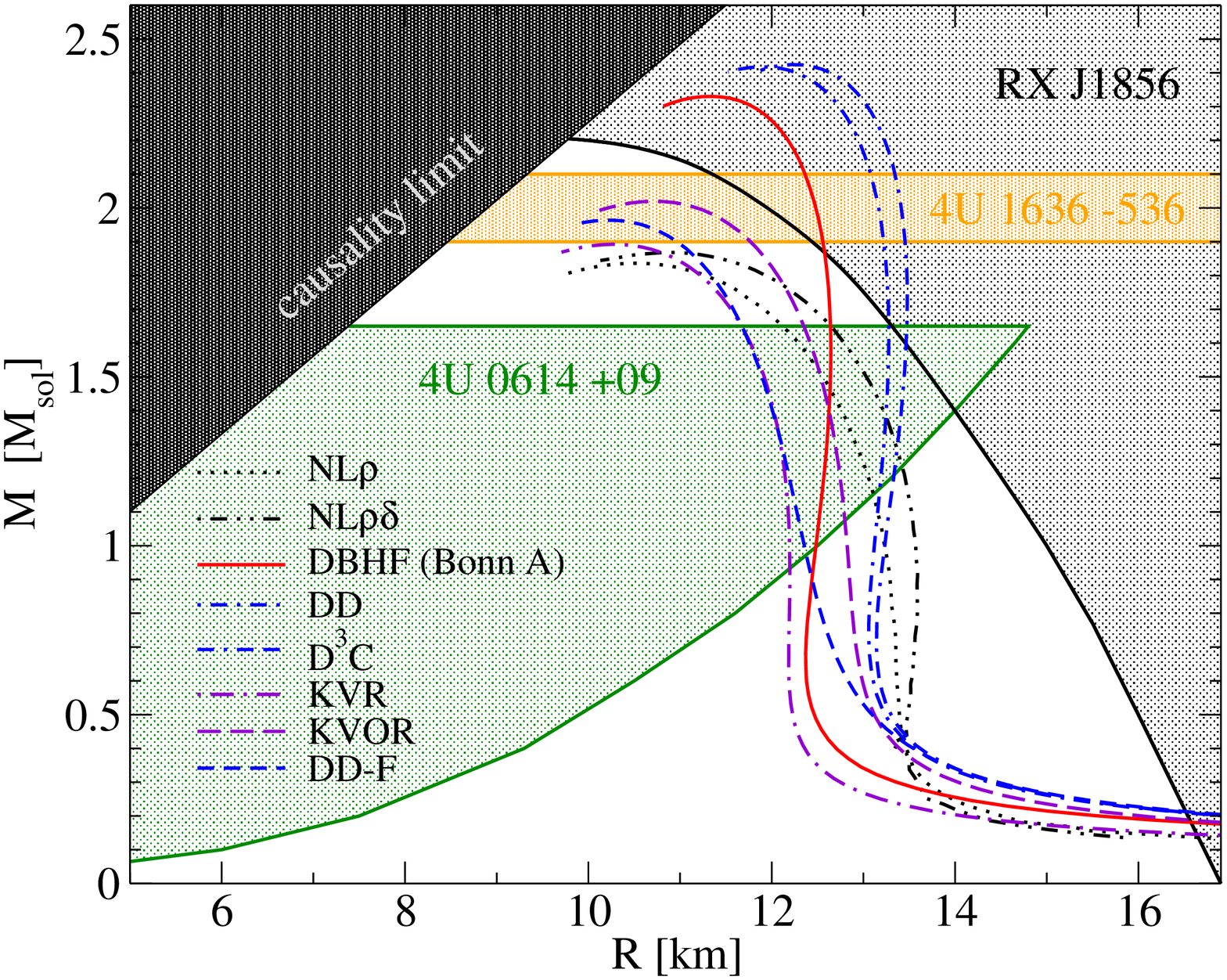}
\hspace{-1cm}\includegraphics[width=7cm,height=7cm,
angle=-90]{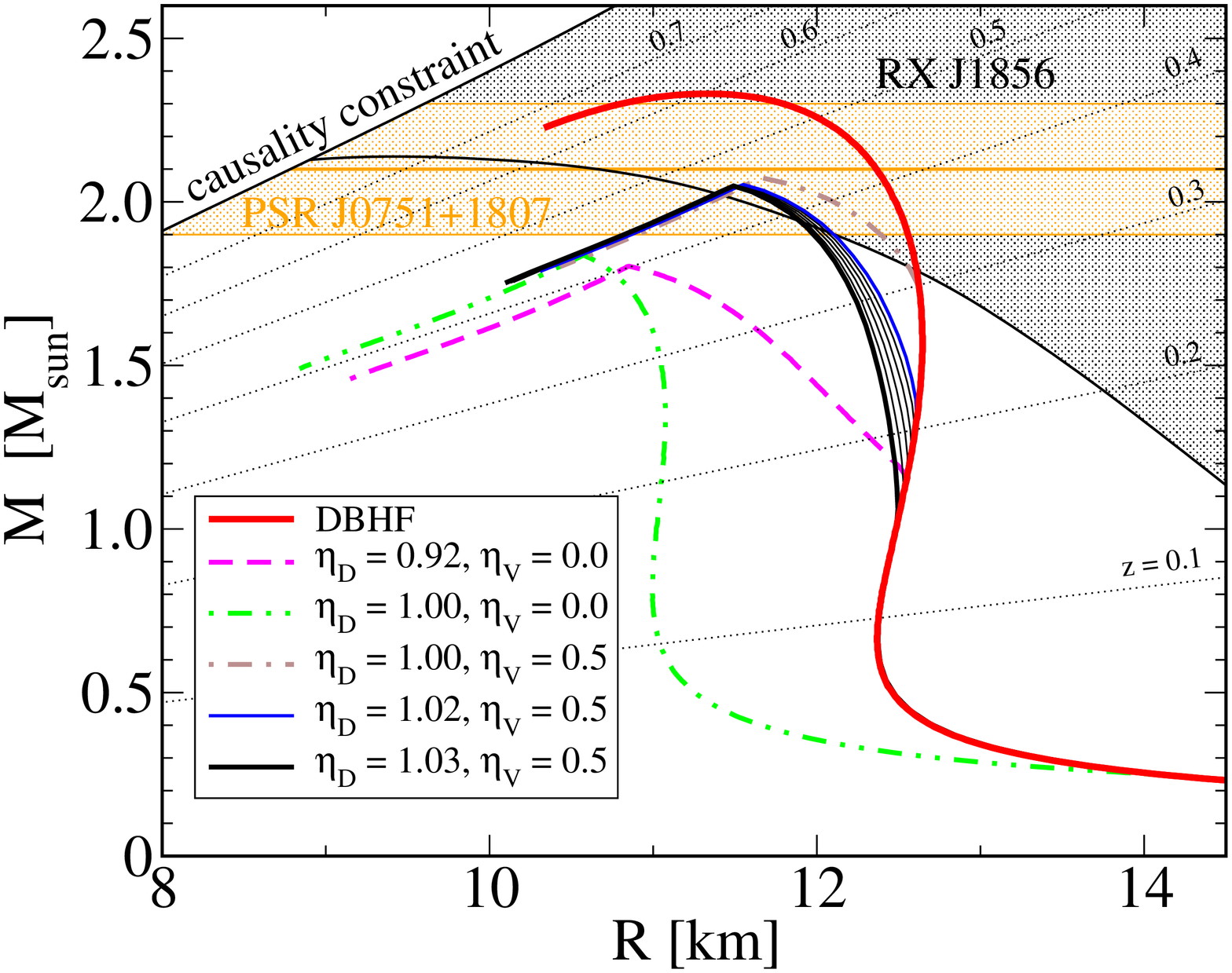}} \caption{ \label{fig:M-R}
Left panel: Mass-Radius constraints from thermal radiation of the
isolated NS RX J1856.5-3754 (grey hatched region) and from QPOs in
the LMXBs 4U 0614+09 (green hatched area) and 4U 1636-536 (orange hatched
region) which shall be regarded as separate conditions to the EsoS.
For the NS in 4U 1636-536 a mass of $2.0\pm 0.1~M_\odot$
is obtained, so that the weak QPO constraint would exclude the
NL$\rho$ and NL$\rho\delta$  EsoS whereas the strong one would
leave only DBHF, DD and D$^3$C.
Right panel: Mass - radius relationship for CS sequences
corresponding to a nuclear matter EoS  (DBHF) and different hybrid
star EsoS (DBHF+NJL), see text and \cite{Klahn:2006iw}.
Indicated are also the constraint
on the mass from the pulsar J0751+1807 \protect{\cite{NiSp05}} and
on the mass-radius relationship from the isolated neutron star RX
J1856  \protect{\cite{Trumper:2003we}}
The dotted lines indicate the gravitational
redshift, $z=(1-2GM/R)^{-1/2}-1$, of photons emitted from the
compact star surface.
Present constraints on the mass-radius relation of CSs do not rule
out hybrid stars. }
\end{figure}

The stellar radius $R$ is defined by zero pressure at the stellar
surface, $P(R)=0$. The star's cumulative gravitational mass is
then given by  $M=m(R)$. Each configuration is specified by the
given central density value $n(0)$ at radius $r=0$.

\subsection{Observational constraint on the Mass and Radius}

On the plot of the Mass-Radius relationship in Fig.~{\ref{fig:M-R}}
we also show regions corresponding to different observed objects.
These results constrain the model EsoS for both hadronic and
hybrid star matter. These data are the following
\begin{itemize}
\item   {\it Mass-Radius relation from LMXBs}\\
as implies \cite{Miller:2003wa,Miller:1996vj} the observation of a
source of quasi-periodic brightness oscillations (QPOs) with a
maximum frequency is $\nu_{\rm max}$ limits the stellar mass and
radius to
\begin{equation}
\begin{array}{rl}
M&<2.2~M_\odot (1000~{\rm Hz}/\nu_{\rm max})(1+0.75j)\\
R&<19.5~{\rm km}(1000~{\rm Hz}/\nu_{\rm max})(1+0.2j)\; ~,
\end{array}
\end{equation}
where $j\equiv cJ/GM^2 \simeq 0.1-0.2 $  is the dimensionless stellar
angular momentum.
  \item {\it Mass-Radius relation from RX J1856:}\\
 The thermal radiation of nearby isolated NS  RX J1856.5-3754
with blackbody spectrum temperature $T_\infty=57$ eV
\cite{Pons:2001px} because of initially estimated distance (60 pc)
has been considered as a self-bound strange quark star with radius
$ R_\infty\approx 8 $ km. For the new measurements (distance $\sim
117$ pc), RX J1856 has to have a rather large radius of $R \sim 14$
km ($R_\infty = 16.8$ km), when the mass is $1.4~M_\odot$
\cite{Trumper:2003we}.
  \item {\it Maximum mass constraint} \\
 from measurements on PSR J0751+1807 imply a pulsar mass of
$2.1\pm0.2\left(^{+0.4}_{-0.5}\right) {\rm M_\odot}$ (first error
estimate with $1\sigma$ confidence, second in brackets with
$2\sigma$ confidence) \cite{NiSp05}.
\end{itemize}

The resulting lower bound in the mass radius plane is shown in
Fig.~{\ref{fig:M-R}}. Particularly interesting is the constraint from
RX J1856 and there are three ways to interpret this result:
\begin{itemize}
\item[A)] RX J1856 belongs to compact stars with typical masses
$M\sim 1.4 M_{\odot}$ and would thus have to have a radius
exceeding $14$ km (see Fig.~\ref{fig:M-R}).
{\em None of the examined EsoS can meet this
requirement. }
\item[B)]  RX J1856 has a typical radius of $R \sim 12 - 13$ km,
implying that the EoS has to be rather stiff at high density in
order to allow for configurations with masses above $\sim
2~M_\odot$. In the present work this condition would be fulfilled
for DBHF, DD and D$^3$C.
This $M>1.7~M_\odot$ explanation implies
that the object is very massive and it is not a typical NS which would
have $M<1.5~M_\odot$, as follows from population synthesis models.
\item[C)] RX J1856 is an exotic object with
a small mass $\sim 0.2~ M_\odot$, which would be possible for all
EsoS considered here.
{\em No such object has been observed yet},
but some mechanisms for their formation and properties have been
discussed in the literature \cite{Popov:2004nw}.
\end{itemize}

\begin{figure}[htb]
\includegraphics[width= 8cm, angle=-90]{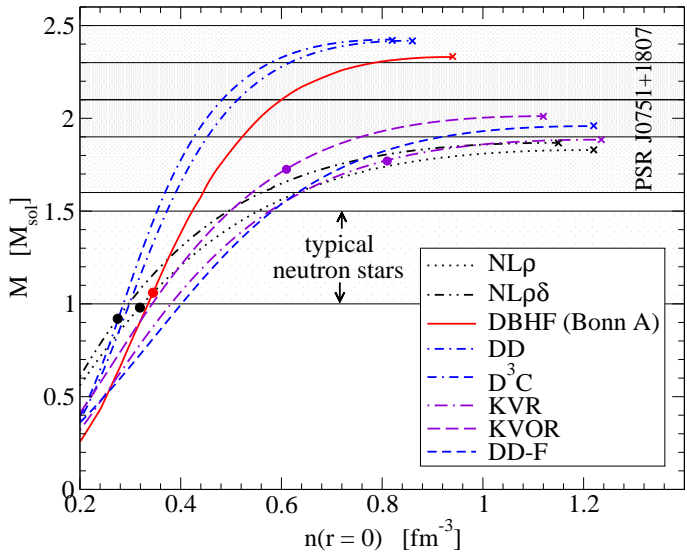}
\caption{\label{fig:m_n}
Mass versus central density for compact star
configurations obtained by solving the TOV equations
(\ref{eq:TOV1}) and (\ref{eq:TOV2}) for all EsoS introduced in
Subsect. 2.1. Crosses denote the maximum mass configurations,
filled dots mark the critical mass and central density values
where the DU cooling process becomes possible. According to the DU
constraint, it should not occur in ``typical NSs'' for which
masses are expected from population synthesis \cite{Popov:2004ey}
to lie in the {lower} grey horizontal band. The dark and light
grey horizontal bands around 2.1 $M_\odot$ denote the 1$\sigma$
and 2$\sigma$ confidence levels, respectively, for the mass
measurement of PSR J0751+1807 \cite{NiSp05}.
 }
\end{figure}

An other constraint on the EoS of stellar matter could be the
critical mass $M_{DU}$ corresponding to the threshold of the most
effective cooling mechanism, so called {\it Direct Urca (DU)
processes} $n\to p+e^-+\bar\nu_e$, which is an ultimate cause of
very fast cooling of stars when its mass is only slightly above the
critical value $M_{DU}$ \cite{BlGrVo04}.
For each hadronic EoS the critical mass is shown in Fig. \ref{fig:m_n}.
It depends on the symmetry
energy $E_S$ and occurs when the proton fraction exceeds $11\%$
(without muons and about $14\%$ with muons in the medium)
\cite{Klahn:2006ir}.

To consider the DU as a strong constraint one needs to investigate
the cooling problem together with the nucleon superfluidity which
suppresses the cooling rates. Calculations show that indeed the
values of the pairing gaps used in the literature are not sufficient
to resolve the DU as a problem \cite{Grigorian:2005fn}.

\section{Cooling behavior of hybrid stars}

The cooling behavior of compact stars belongs to the most complex
phenomena in astrophysics. Therefore, the codes for its numerical
simulation as developed by a few groups contain inputs (cooling
regulators) of rather different kind, see, e.g.,
\cite{Page:2005fq,Page:2004fy,Blaschke:2004vq,Voskresensky:2001fd,Yakovlev:2000jp}.
Attempts to develop a {\em Minimal Cooling Paradigm}
\cite{Page:2004fy} by omitting important medium effects on cooling
regulators \cite{Voskresensky:2001fd,Grigorian:2005fn}
unfortunately result in inconsistencies and suffer therefore from
the danger of being not reliable. To develop a paradigmatic
cooling code as an open standard, however, is rather necessary to
cross-check the present knowledge of the groups before more
sophisticated mechanisms like anisotropies due to the magnetic
field \cite{Page:2005fq} or special processes in the NS crust or
at the surface are taken into account. Therefore, it is still
premature to attempt an identification of the NS interior from the
cooling behavior.

In order to circumvent such a model dependence we employ a given
cooling code developed in Refs.
\cite{Blaschke:2004vq,Blaschke:2000dy} and vary the matter
properties such as EoS, superconductivity and star crust model
such as to fulfill all constraints known up to now (mass,
mass-radius, TA, brightness, etc.). Moreover, we try to use
consistent inputs.

The main neutrino cooling processes  in hadronic matter are the
direct Urca (DU), the medium modified Urca (MMU) and the pair
breaking and formation (PBF) whereas in quark matter the main
processes are the quark direct Urca (QDU), quark modified Urca
(QMU), quark bremsstrahlung (QB) and quark pair formation and
breaking (QPFB) \cite{Jaikumar:2001hq}. Also the electron
bremsstrahlung (EB), and the massive gluon-photon decay (see
\cite{Blaschke:1999qx}) are included.

The $1S_0$ neutron and proton gaps in the hadronic shell are taken
according to the calculations by \cite{Takatsuka:2004zq}
corresponding to the thick lines in Fig. 5 of  Ref.
\cite{Blaschke:2004vq}. However, the $3P_2$ gap is suppressed by a
factor 10 compared to the BCS model calculation of
\cite{Takatsuka:2004zq}, consistent with arguments from a
renormalization group treatment of nuclear pairing
\cite{Schwenk:2003bc}. Without such a suppression of the $3P_2$
gap the hadronic cooling scenario would not fulfill the TA
constraint, see \cite{Grigorian:2005fn}.

\begin{center}
\begin{figure}[th]
\centerline{
\hspace{1cm}\psfig{figure=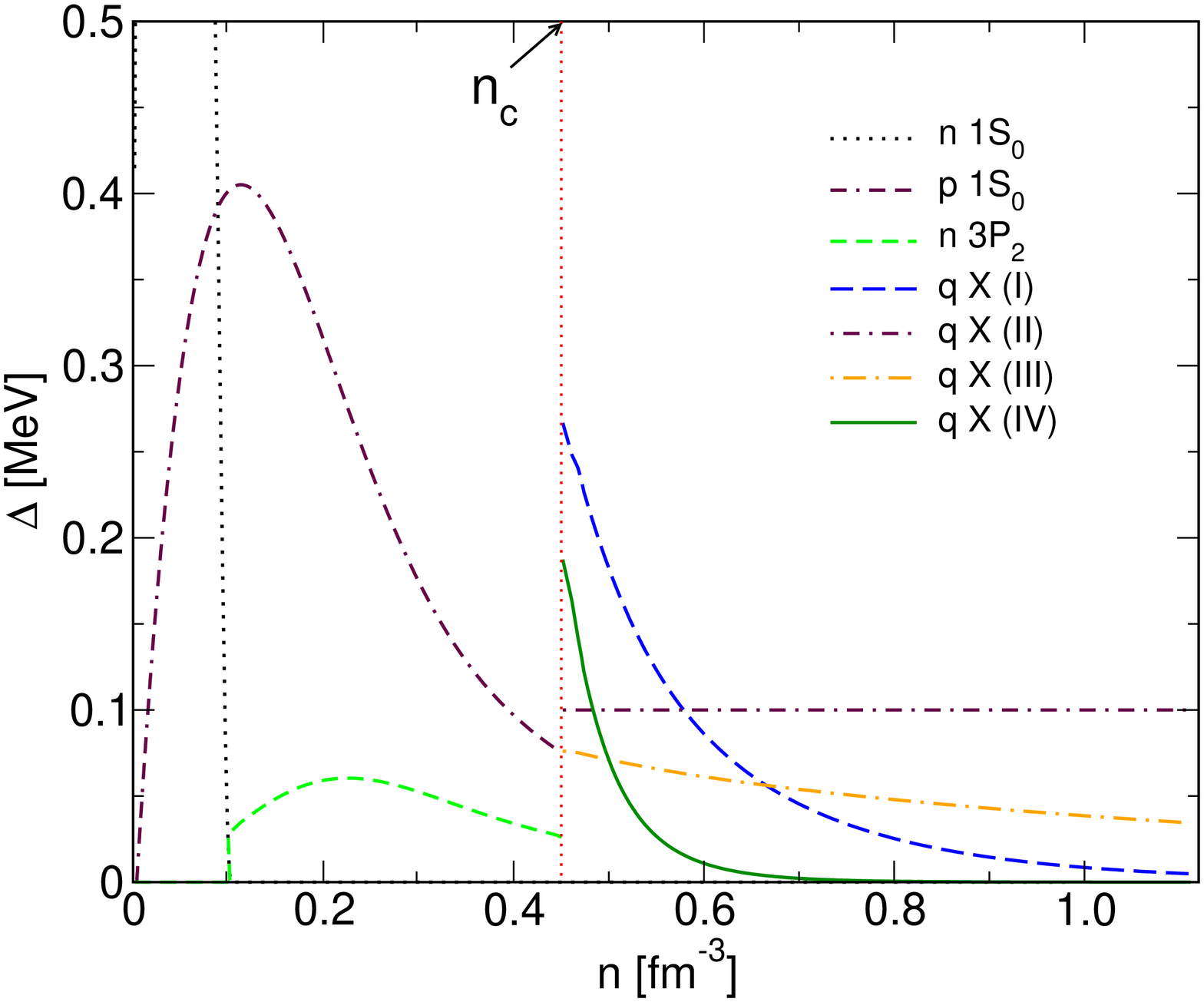,height=7cm,width=6cm,angle=-90}
\hspace{-5mm}\psfig{figure=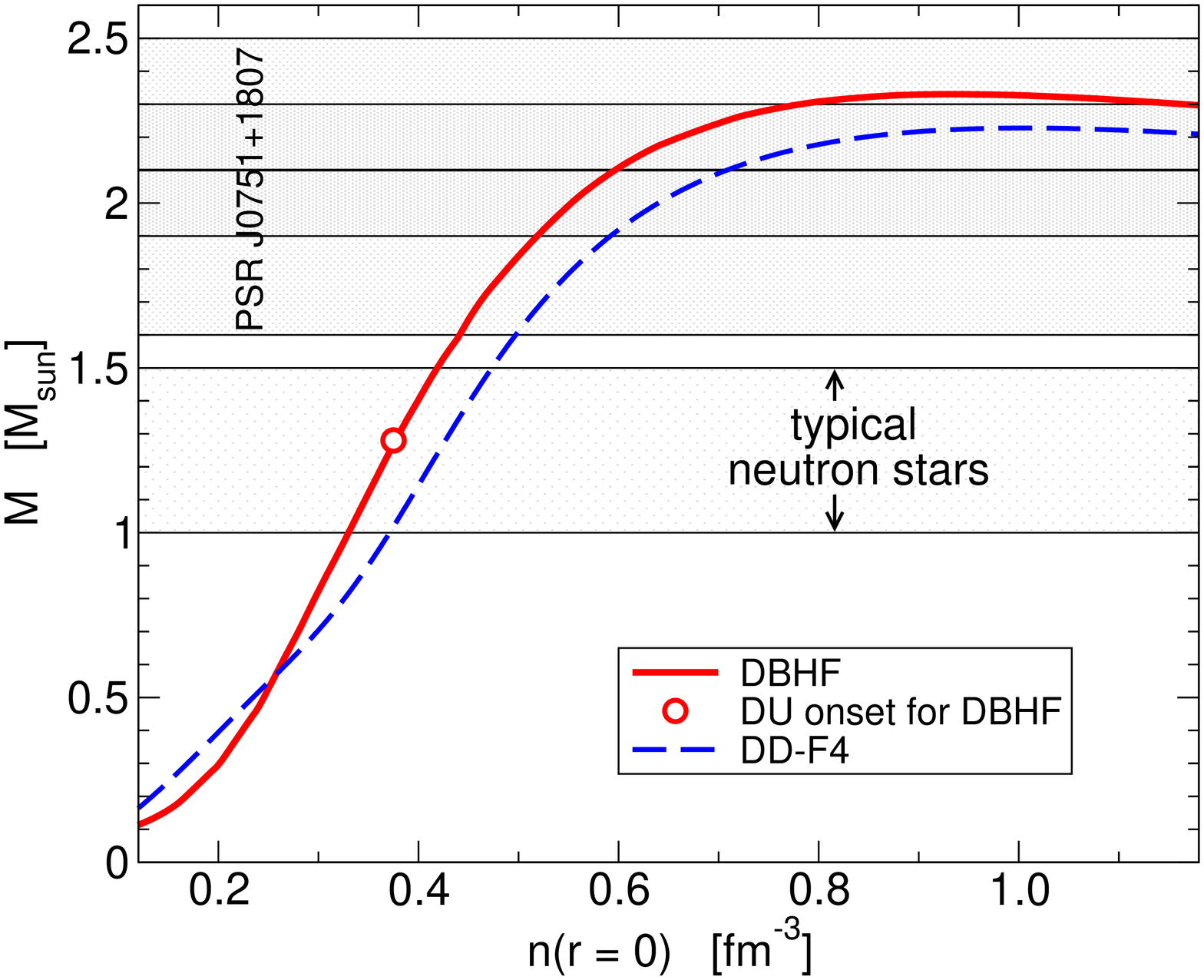,height=7cm,width=6cm,angle=-90}}
\caption{{\small Density dependence of the pairing gaps in nuclear
matter together with that of the hypothetical X-gap in quark
matter (left). Mass-central density relation for the two hadronic
EoS models DBHF and DD-F4 (right). The dot indicates the onset of
the DU process.} \label{fig:gaps}}
\end{figure}
\end{center}

The possibilities of pion condensation and of other so called
exotic processes are included in the calculations for purely
hadronic stars but do not occur in the hybrid ones since the
critical density for pion condensation exceeds that for
deconfinement in our case \cite{Blaschke:2004vq}. While the
hadronic DU process occurs in the DBHF model EoS for all neutron
stars with masses above $1.27~M_\odot$, it is not present at all
in the DD-F4 model, see the right panel of Fig.~\ref{fig:gaps}. We
account for the specific heat and the heat conductivity of all
existing particle species contributing with fractions determined
by the $\beta-$ equilibrium conditions. Additionally, in quark
matter the massless and massive gluon-photon modes also
contribute.

In the 2SC phase only the contributions of quarks forming Cooper
pairs (say red and green) are suppressed via huge diquark gaps,
while those of the remaining unpaired blue color lead to a so fast
cooling that the hybrid cooling scenario becomes unfavorable
\cite{Grigorian:2004jq}. Therefore, we assume the existence of a
weak pairing channel such that in the dispersion relation of
hitherto unpaired blue quarks a small residual gap can appear. We
call this gap $\Delta_X$ and show that for a successful
description of the cooling scenario $\Delta_X$ has to have a
density dependence. We have studied the ansatz
$\Delta_{\mathrm{X}}= \Delta_0 \, \exp{\left[-\alpha\, (\mu/ \mu_c
- 1)\right]} $, where $\mu$ is the quark chemical potential,
$\mu_c=330$ MeV. For the analyses of possible models we vary the
values of $\alpha$ and $\Delta_0$, given in the Table 1 of
\cite{Popov:2005xa} and shown in the left panel of Fig.
\ref{fig:gaps}.

The physical origin of the X-gap is not yet identified. It could
occur, e.g., due to quantum fluctuations of color neutral quark
{sextet} complexes \cite{Barrois:1977xd}. Such calculations have
not yet been performed within the relativistic chiral quark
models. The size of the small pairing gaps in possible residual
single color/ single flavor channels \cite{Schafer:2000tw} is
typically in the interval $10~$ keV - $1~$MeV, see discussion in
\cite{Alford:2002rz}. The specific example of the CSL phase is
analyzed in  more in detail in Refs.
\cite{Aguilera:2005tg,Schmitt:2005wg,Aguilera:2005uf,Aguilera:2006cj}.

\subsection{Cooling curves in the TA diagram}

We consider the cooling evolution of young neutron stars with ages
$t \sim 10^3 - 10^6$ yr which is governed by the emission of
neutrinos from the interior for $t \lsim 10^5$ yr and thermal
photon emission for $t \gsim 10^5$ yr. The internal temperature is
of the order of $T \sim 1$ keV. This is much smaller than the
neutrino opacity temperature $T_{\rm opac} \sim 1$ MeV as well as
critical temperatures for superconductivity  in nuclear ($T_c \sim
1$ MeV) or quark matter ($T_c \sim 1 - 100$ MeV). Therefore, the
neutrinos are not trapped and the matter is in a superconducting
state. In Fig. \ref{fig:gaps} we show the density dependence of
the pairing gaps in nuclear matter
\cite{Takatsuka:2004zq,Blaschke:2004vq} together with that of the
hypothetical X-gap in quark matter
\cite{Blaschke:2004vr,Grigorian:2004jq,Popov:2005xa}. The phase
transition occurs at the critical density $n_c = 2.75~n_0=0.44$
fm$^{-3}$.

\begin{center}
\begin{figure}[th]
\centerline{
\hspace{1cm}\psfig{figure=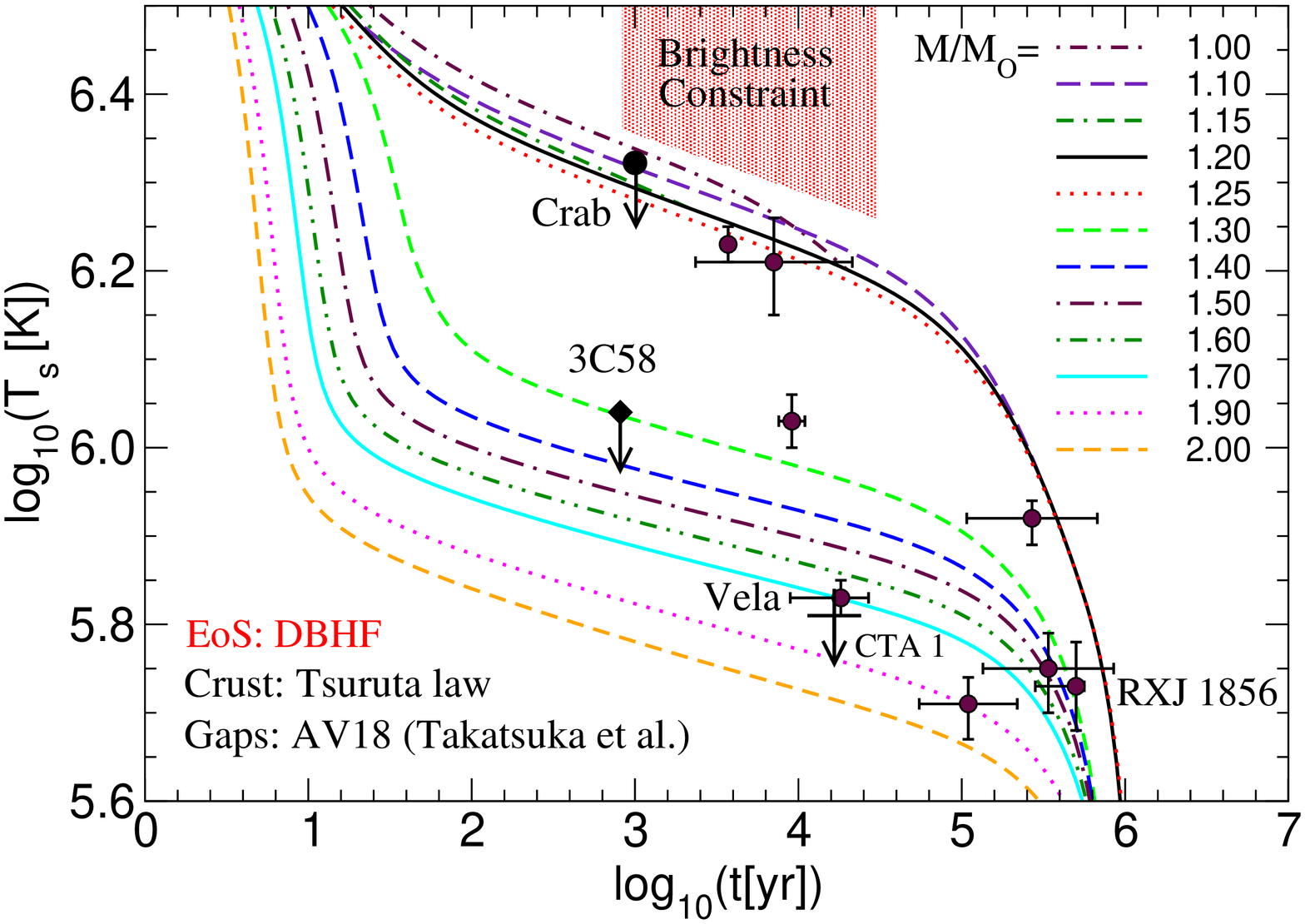,height=7cm,width=8cm,angle=-90}
\hspace{-4mm}\psfig{figure=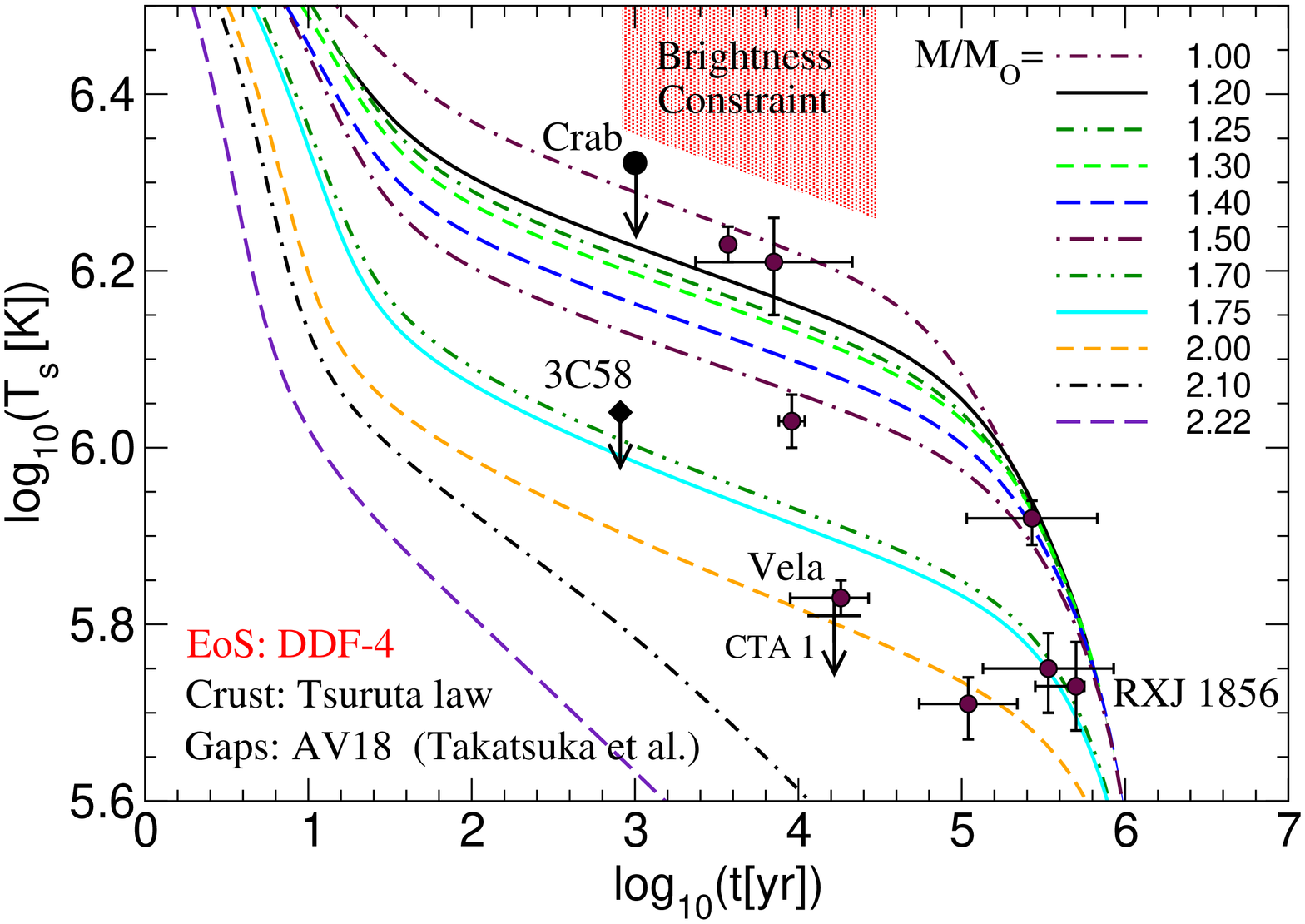,height=7cm,width=8cm,angle=-90}}
\caption{{\small Hadronic star cooling curves for DBHF model EoS.
Different lines correspond to compact star mass values indicated
in the legend (in units of $M_\odot$), data points with error bars
are taken from Ref. \cite{Page:2004fy}.} \label{fig:hc1}}
\end{figure}
\end{center}

\begin{center}
\begin{figure}[th]
\centerline{
\hspace{0cm}\psfig{figure=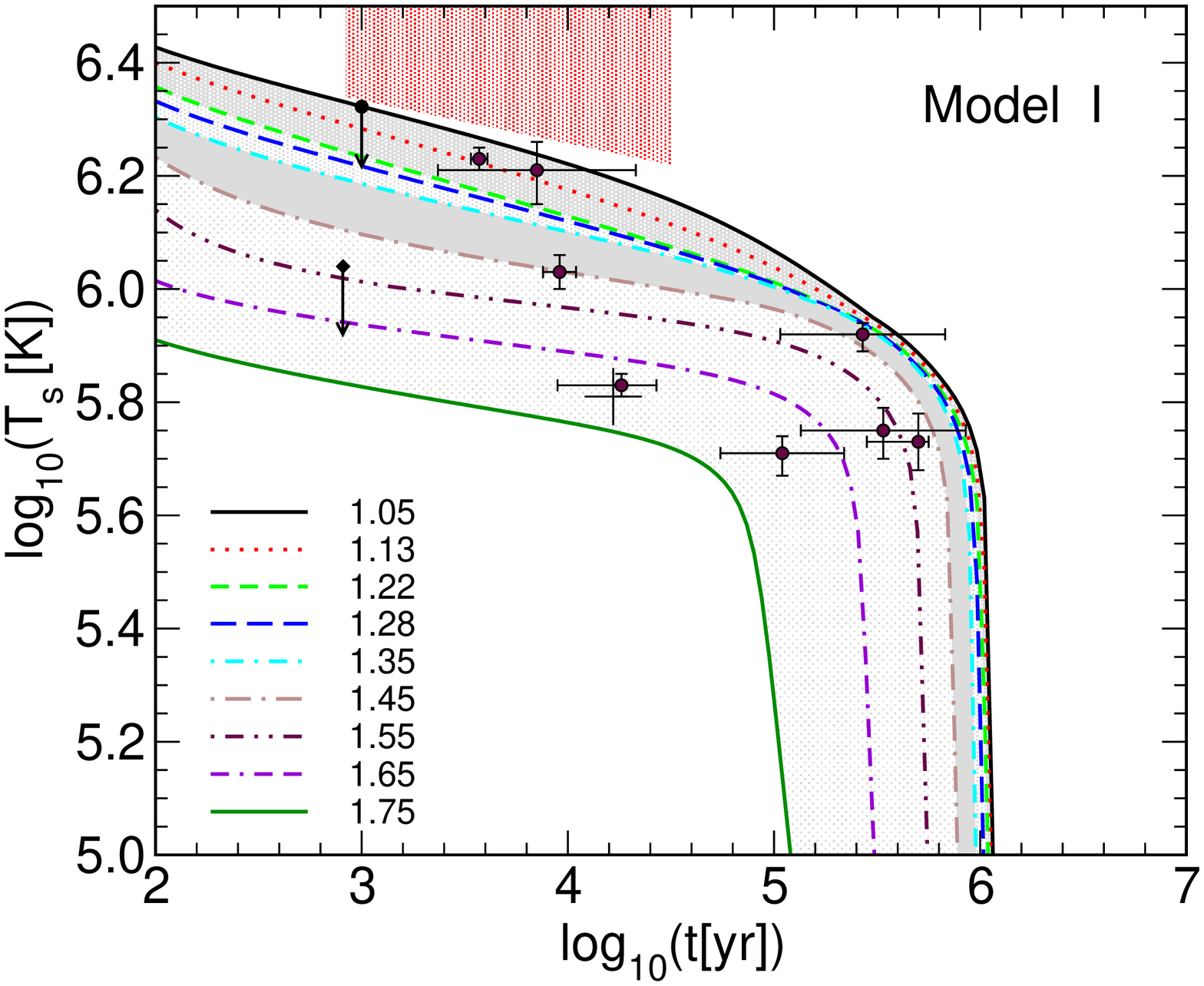,height=7cm,width=8cm,angle=-90}
\hspace{-1cm}\psfig{figure=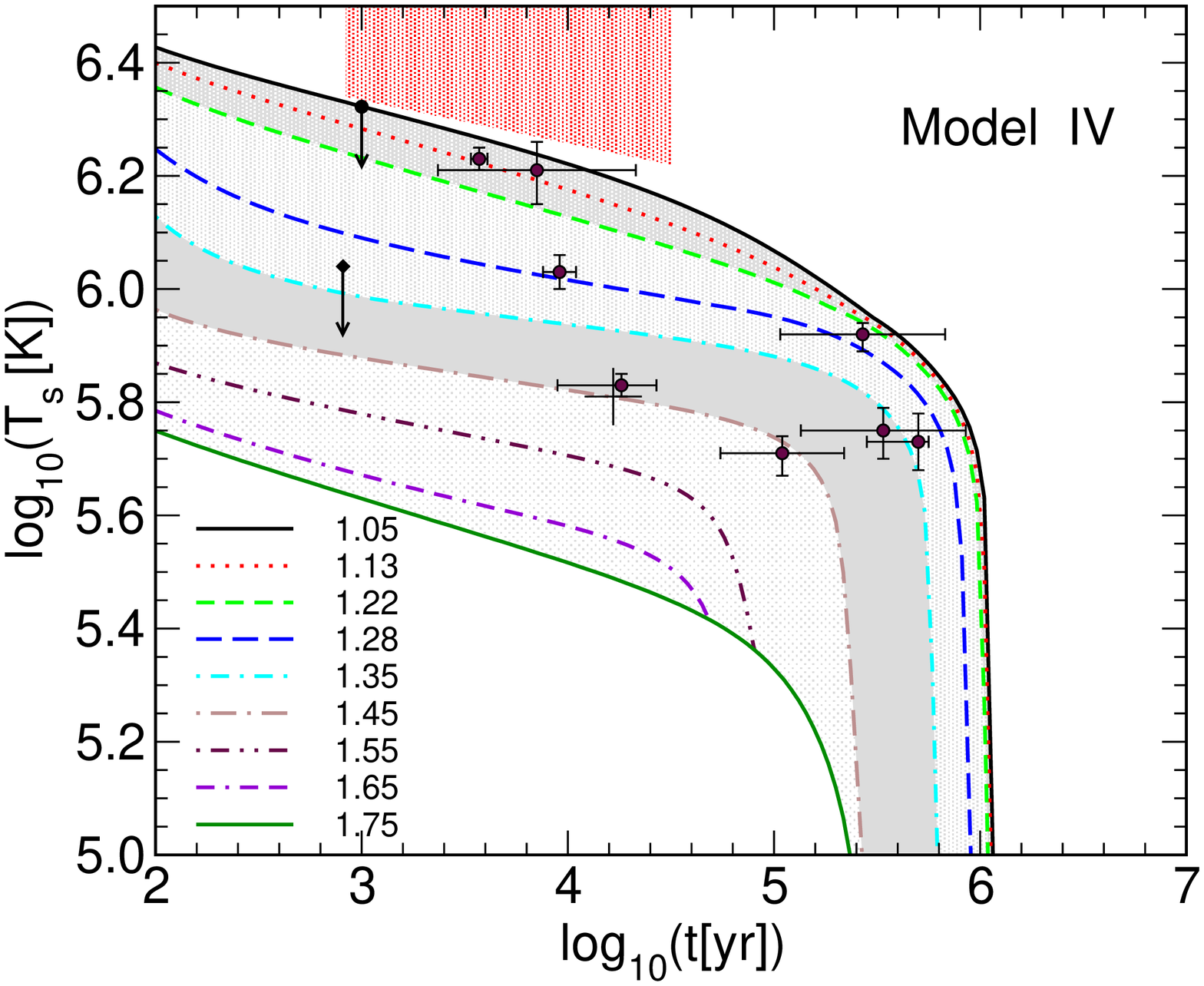,height=7cm,width=8cm,angle=-90}}
\caption{{\small Cooling curves for hybrid star configurations
with 2SC+X pairing pattern and X-gap model I (left) versus model
IV (right). For the gaps see the left panel of Fig.~1. The grey
value for the shading of the mass bin areas corresponds to the
probability for that mass bin value in the population synthesis
model of Ref. \cite{Popov:2004ey}.} \label{fig:qc1}}
\end{figure}
\end{center}

In Fig.~\ref{fig:hc1} we present TA-diagrams for two different
hadronic models and in Fig.~\ref{fig:qc1} for two hybrid star cooling models
presented in Ref. \cite{Popov:2005xa}. In the cooling calculations
presented the crust model, i.e. the $T_m-T_s$ relationship between be
temperatures of the inner crust and the surface \cite{Blaschke:2004vq},
has been chosen as such to fulfill the TA test: each data point should be
explained with cooling curve belonging to an admissible configuration.

The TA data points are taken from \cite{Page:2004fy}. The hatched
trapeze-like region represents the brightness constraint (BC)
\cite{Grigorian:2005fd}. For each model nine cooling curves are
shown for configurations with mass values corresponding to the
binning of the population synthesis calculations explained in
\cite{Popov:2005xa}.

In \cite{Popov:2005xa} for these hybrid cooling scenario the
logN-LogS distribution constraint has been considered and it has
been suggested to use the marking of TA diagram with five grey
values in order to encode the likelihood that stars in that mass
interval can be found in the solar neighborhood, in accordance
with the population synthesis scenario, see Fig. \ref{fig:qc1}.
The darkest grey value, for example, corresponds to the most
populated mass interval $1.35$ - $ 1.45~M_\odot$ predicted by the
mass spectrum used in population synthesis.


On the TA- diagram we show also the other constraint given by the
{\it maximum brightness} of CSs, as discussed by
\cite{Grigorian:2005fd}. It is based on the fact that despite many
observational efforts one has not observed very hot NSs ($\log$T
$> 6.3-6.4$ K) with ages of $10^3$ - $10^{4.5}$ years. Since it
would be very easy to find them - if they exist in the galaxy -
one has to conclude that at least their fraction is very small.
Therefore a realistic model  should not predict CSs with typical
masses at temperatures noticeable higher than the observed ones.
The region of avoidance is the hatched trapezoidal region in
Figures \ref{fig:hc1} and \ref{fig:qc1}.

Altogether, the hybrid star cooling behavior obtained for our EoS
fits all of the sketched constraints under the assumption of the
existence of a 2SC phase  with X-gaps.

\section{Unmasking neutron star interiors using cooling simulations}

Considering the interior of neutron stars as a ''laboratory''
where matter under conditions of extreme densities occurs
\cite{Weber:1999qn} we hope that it is possible to identify the
composition of neutron star interiors from their observable
properties like, e.g., masses, radii, rotational and cooling
evolution \cite{Prakash:1996xs,Blaschke:2001uj}.

The question for the existence of quark matter inside a compact object
has often been mismatched with that of the existence of strange stars,
essentially made up of strange quark matter without a hadronic shell and
thus very compact with radii less than $\sim 10$ km
\cite{Bombaci:2001uk,Li:1999wt}.
The measurements of high masses for objects like the pulsar PSR J0751+1807
\cite{NiSp05} with $M=2.1 \pm 0.2~M_\odot$ or of the mass-radius relation
from the RX J1856-3754 pointing to either large masses or large radii of
$R> 14$ km \cite{Trumper:2003we} require a stiff equation of state and
exclude standard models for hyperonic or quark matter interiors as
well as mesonic condensates, see \cite{Ozel:2006bv}.

Therefore, all suggested signals which are based on large first order
phase transition effects and changes in the mechanical roperties such as
the timing behavior of pulsar spin-down \cite{Glendenning:1997fy},
frequency clustering \cite{Glendenning:2000zz} or population clustering
\cite{Poghosyan:2000mr,Blaschke:2001th} of accreting NSs should
not be applicable.

Here we discuss a new, sensitive tool for ''unmasking'' the composition
of neutron stars which is based on their cooling behavior
\cite{Blaschke:2006gd}.
As the cooling regulators such as neutrino emissivities, heat
conductivity and specific heat in quark matter might be
qualitatively different from those in nuclear matter, due to the
chiral transition and color superconductivity with some possibly
sensible density dependence, the TA curves for hybrid stars could
be significantly different from those of neutron stars.


\subsection{Mass distribution from TA data}

In order to reach the goal of unmasking the neutron star interior
we use here a new method for the quantitative analysis of
the cooling behavior consisting in the extraction of a NS mass
distribution from the (yet sparse) TA data \cite{Blaschke:2006gd}
and its comparison with
the (most likely) mass distribution from population synthesis
models of NS evolution in the galaxy \cite{Popov:2004ey}.

This method can be described as follows.
For a given cooling model defined by the EoS and the cooling regulators,
to each configuration with a gravitational mass $M$ corresponding
cooling curve $T(t;M)$ can be determined. For each set of mass
values $M_i$, $i=0,\dots,N_M$ which is defining the borders of
$N_M$ mass bins as, for example, in the population synthesis one
can mark a region with borders of pair of neighboring cooling
curves $T(t;M_i)$ and $T(t;M_{i-1})$. This strip in the TA plane
corresponds to the $i^{th}$ mass bin. As a measure for the number
of cooling objects to be expected within this mass bin we chose
\begin{equation}
N_i=\sum_{j=1}^{N_{\rm cool}}\int dt\int_{T(t;M_i)}^{T(t;M_{i-1})}
dT P_j(T,t), \label{Ni}
\end{equation}
where $N_{\rm cool}$ denotes the total number of observed coolers
used for the analysis and $P_j(T,t)$ is the probability density to
find the $j^{th}$ object at the point $(T,t)$ in the TA plane.
In Ref. \cite{Blaschke:2006gd} the simplest ansatz has been made that
$P_j(T,t)$ is constant in the rectangular region defined by the
upper and lower limits of the confidence intervals corresponding
to the temperature and age measurements, $(T_{jl},T_{ju})$ and
$(t_{jl},t_{ju})$, respectively,
\begin{equation}
P_j(T,t)=[(t_{jl}-t_{ju})(T_{jl}-T_{ju})]^{-1}\Theta(T-T_{jl})\Theta(T_{ju}-T)
\Theta(t-t_{jl})\Theta(t_{ju}-t)~. \label{Pj}
\end{equation}
Note that in the case when the exact age the object is known
(e.g., for a historical supernova), the time-dependence of
$P_j(T,t)$ degenerates to a $\delta$-function and the $t$-integral
in (\ref{Ni}) can be immediately carried out, leaving us with a
one-dimensional probability measure.

\begin{center}
\begin{figure}[h!]
\centerline{
\hspace{1cm}\psfig{figure=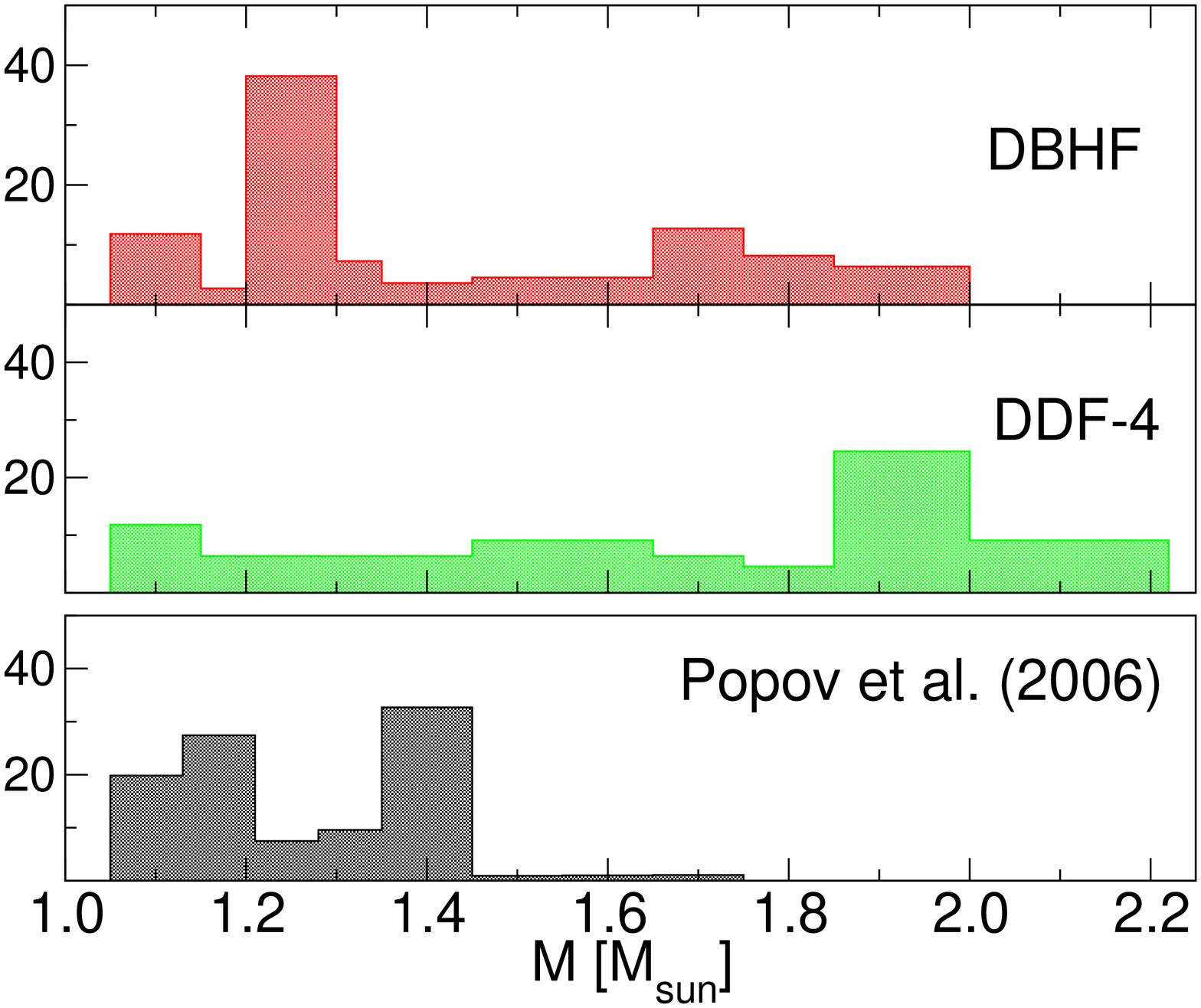,height=7cm,width=8cm,angle=-90}
\hspace{-5mm}\psfig{figure=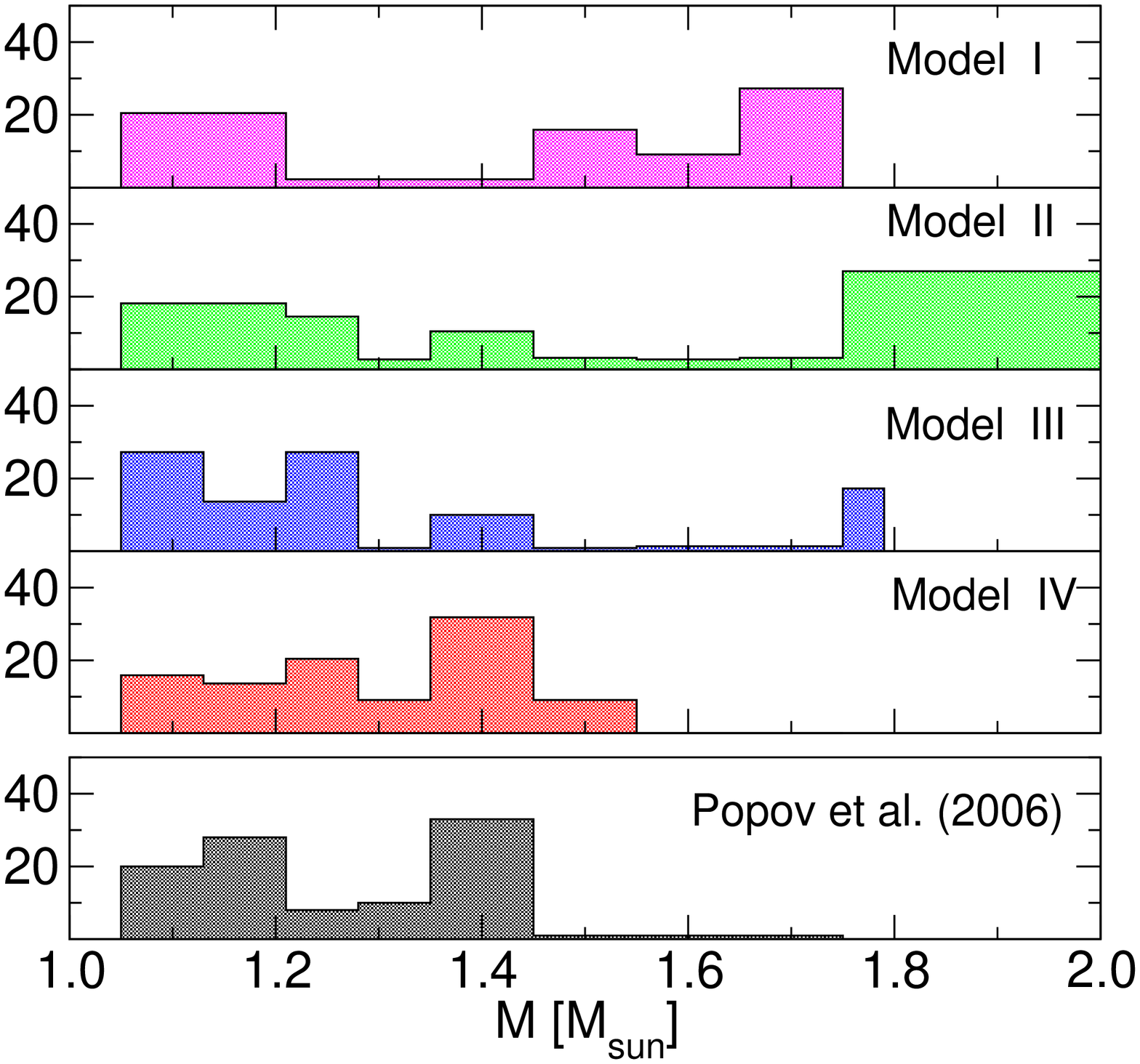,height=7cm,width=8cm,angle=-90}}
\caption{{\small NS mass spectra extracted from the distribution
of cooling data for both hadronic EoS models (left panel) and for
hybrid stars with X-gap models I-IV (right panel). For comparison,
the mass disctribution of young, nearby NS from the population
synthesis of Popov et al. \cite{Popov:2005xa} is shown at the
bottom of the panels. } \label{fig:masshb}}
\end{figure}
\end{center}

This method has been applied to the cooling models for hadronic
and hybrid stars described in the previous section. The results
for the extracted mass distributions are normalized to 100
objects, defining $N(M)=100 ~ N_i/(\sum_{i=0}^{N_M} N_i)$, and
shown in Fig.~\ref{fig:masshb}.

As we see from Fig.~\ref{fig:masshb}, the results are very sensitive to the
chosen cooling model. In the hadronic scenario the onset of the DU
cooling mechanism drastically narrows the mass distribution around
the critical mass for the DU onset, see Fig.~\ref{fig:hc1}. On the other hand
the slow cooling model predicts more massive objects than could be
justified from the independent population analysis.

When comparing the density dependence of the pairing gaps, given
in the left panel of Fig.~\ref{fig:gaps}, with the extracted mass
distributions for the corresponding hybrid models in the right panel of
Fig.~\ref{fig:masshb},
the direct relationship between the superconductivity and the mass
distribution becomes obvious.

The DU problem as it was previously discussed in the literature
\cite{Blaschke:2004vq,Klahn:2006ir,Kolomeitsev:2004ff} was based
on the intuitive understanding that the mass distribution can not
be peaked at a critical mass value which accidentally is unique
for all observed young objects. Our modification of the definition
of the DU problem does not contradict that suggestion, but rather
provides a quantitative measure which to some extent rehabilitates
the validity of cooling scenarios including the DU process.

On the other hand, the EoS model should obey the mass constraints
too. Therefore, using the models discussed in this work we
demonstrate that the most preferable structure of the compact
object is likely to be a hybrid star with  properly defined color
superconductivity of the quark matter state in the core.

\section{Conclusions}

\label{sec:firstconcl}
In these lectures we have discussed the recent
developments towards a test scheme for the EoS of matter under the extreme
conditions using observational data from
compact stars and experimental results from heavy-ion collisions.

The application of this scheme to specific EsoS offers some
interesting insights which reveal the discriminative power of
their combined tests in a broad region of densities including the
possibility of a deconfinement phase transition.

In addition to the astrophysical part of the tests the results for
the elliptic flow \cite{DaLaLy02} restrict the hardness of
symmetric matter which is in opposition to the constraint from
neutron star maximum mass and/or mass-radius measurements.

One of the main results concludes that no present phenomenological
finding bears a strong argument against the presence of a QM core
inside NSs \cite{Alford:2006vz,Klahn:2006iw}.

The resulting phase diagram for stellar as well as symmetric matter
including the transition from nuclear to quark matter and satisfying the
constraints discussed here exhibits almost a crossover transition with a
negligibly small coexistence region and a tiny density jump.
Therefore, the differences between the mechanical properties of the hadronic
and hybrid stars are very tiny and the problem of the
cooling history of the compact stars becomes central for the
discrimination between alternatives for the composition of their interiors.

We have presented mass distributions obtained by analysing cooling
calculations for NSs and demonstrated that they provide a sensible
measure for the composition of compact star matter, even if the
mechanical properties of the compact objects are almost identical.

\section*{Acknowledgements}
I thank the organizers of the Helmholtz International Summer School on
"Dense Matter in Heavy-Ion Collisions and Astrophysics" at JINR Dubna
providing me the opportunity to participate in the school. I am grateful
to D. Blaschke for useful comments and discussions. I also acknowledge
all my coauthors for our collaboration.


\begin{thebibliography}{99}
\bibitem{Klahn:2006ir}
  {\it ~Kl\"ahn T. et al.} //
  Phys. Rev. C. 2006. V.74. P.035802.

\bibitem{Klahn:2006iw}
  {\it ~Kl\"ahn T. et al.} //
  arXiv:nucl-th/0609067.

\bibitem{Alford:2006vz}
\textit{~Alford M. et al.} //
Nature 2007. {V.445}, P.E7.

\bibitem{Alford:2004pf}
\textit{~Alford M.,~Braby M.,~Paris M.~W. and~Reddy S.} //
\APJ 2005. V.629. P.969.

\bibitem{Lattimer:2000nx}
\textit{~Lattimer  J.~M. and~Prakash M.}//
  Astrophys. J. 2001. V.550. P.426.

\bibitem{BaBu00}
\textit{~Baldo M.,~Burgio G.~F.,~Schulze H.~J.}//
 Phys. Rev. C. 2000. V.61. P.055801.

\bibitem{Glendenning:1992vb}
  \textit{~Glendenning N.~K.}//
  Phys.\ Rev.\ D. 1992. {V.46}. P.1274.

\bibitem{Voskresensky:2002hu}
\textit{~Voskresensky D.~N.,~Yasuhira M. and~Tatsumi T.} //
  Nucl.\ Phys.\ A. 2003 V.723. P.291.

\bibitem{Maruyama:2005tb}
\textit{~Maruyama T.,~Tatsumi T.,~Voskresensky D.~N.,~Tanigawa T.,
~Endo T.and~Chiba S.} //
  arXiv:nucl-th/0505063.

\bibitem{Wal74}
 \textit{Walecka J.} //
 Ann. Phys. (N.Y.) 1974. V.83. P.491;\\
 \textit{Serot B. D. and Walecka J. D.}//
 Adv. Nucl. Phys. 1986. V.16. P.1.

\bibitem{Tok98}
 \textit{Toki H. et al.}//
 J. Phys. G. 1998. V.24 P.1479.

\bibitem{Lal05} 
 \textit{Lalazissis G. A., Nik$\check{\rm s}$i\'{c} T., Vretenar D. and Ring P.}//
 Phys. Rev. C. 2005. V.71. P.024312.

\bibitem{Typel:2005ba}
  \textit{~Typel S.} //
  Phys. Rev. C. 2005. V.71. P.064301.

\bibitem{honnef}
\textit{Fuchs C.}
// Lect. Notes Phys. 2004. V.641. P.119.

\bibitem{DaFuFae05}
\textit{~van~Dalen E. N. E.,~Fuchs C. and~Faessler A.}// Nucl.
Phys. A. 2004. V744.  P.227;
Phys. Rev. C. 2005. V.72. P.065803; 
Phys. Rev. Lett. 2005. V.95. P.022302.


\bibitem{KoVo05}
        \textit{~Kolomeitsev E. E. and~Voskresensky  D. N.} //
        Nucl. Phys. A. 2005. V.759. P.373.

\bibitem{AkPaRa98}
        \textit{Akmal A., Pandharipande V. R., Ravenhall D. G.} //
        Phys. Rev. C 1998. V.58. P.1804.


\bibitem{Brown:1991kk}
\textit{~Brown G.~E. and~Rho M.} //
  Phys.\ Rev.\ Lett.\  1991. V.66. P.2720.

\bibitem{Buballa:2003qv}
  \textit{~Buballa M.} //
Phys. Rep. 2005. {V.407. }{P.205.}

\bibitem{Blaschke:2005uj}
\textit{~Blaschke D.,~Fredriksson S.,~Grigorian H.,~\"Oztas A.~M.
and~Sandin F.} //
Phys.\ Rev.\ D. 2005. {V.72.}{ P.065020.}

\bibitem{Ruster:2005jc}
\textit{~R\"uster S.~B.,~Werth V.,~Buballa M.,~Shovkovy I.~A.
and~Rischke D.~H.} //
  Phys.\ Rev.\ D. 2005. V.72. P.034004.

\bibitem{Abuki:2005ms}
 \textit{~Abuki H. and~Kunihiro T.} //
  Nucl.\ Phys.\ A. 2006. V.768. P.118.


\bibitem{Kapusta:2006}
\textit{ Kapusta~J. and Gale~C.} // {\it Finite temperature Field
Theory}, Cambridge University Press, Cambridge (2006).

\bibitem{Grigorian:2006qe}
  \textit{~Grigorian H.} //
  Phys. Part. Nucl. Lett. 2007. V.4. P.382.

\bibitem{Grigorian:2006pu}
  \textit{~Grigorian H.,~Blaschke D. and~Kl\"ahn T.} //
\Journal{\arx}{astro-ph/0611595}{}{2006}

\bibitem{DaLaLy02} 
 \textit{~Danielewicz P.,~Lacey R. and~Lynch W.~G.} //
  Science 2002. V.298. P.1592.

\bibitem{BaPe71}
\textit{~Baym G.,~Pethick C.,~Sutherland P.} //
 Astrophys. J. 1971. V.170. P.299.

\bibitem{NiSp05}
        \textit{~Nice D. J. et al.} //
        Astrophys. J. 2005. V.634. P.1242.

\bibitem{Trumper:2003we}
  \textit{~Tr\"umper J.~E.,~Burwitz V.,~Haberl F. and~Zavlin
  V.~E.} //
  Nucl.\ Phys.\ Proc.\ Suppl.\ 2004. V.132. P.560.

\bibitem{Miller:2003wa}
  \textit{~Miller M.~C.} //
  AIP Conf.\ Proc.\ 2004. V.714. P.365.

\bibitem{Miller:1996vj}
 \textit{~Miller M.~C.,~Lamb F.~K. and~Psaltis D.} //
  Astrophys.\ J.\  1998. V.508. P.791.

\bibitem{Pons:2001px}
\textit{~Pons J.~A. et al.} //
  Astrophys.\ J.\ 2002. V.564. P.981.

\bibitem{Popov:2004nw}
  \textit{~Popov S.~B.} //
  arXiv:astro-ph/0403710.

\bibitem{BlGrVo04}
\textit{~Blaschke D.,~Grigorian H. and~Voskresensky D.} //
 Astron. Astrophys. 2004. V.424. P.979.


\bibitem{Grigorian:2005fn}
\textit{~Grigorian H. and~Voskresensky D.~N.} //
 Astron. Astrophys. 2005. {V.444.}{ P.913.}


\bibitem{Popov:2004ey}
  \textit{~Popov S.,~Grigorian H.,~Turolla R. and~Blaschke D.}
\Journal{\AA}{448}{327}{2006}

\bibitem{Page:2005fq}
  \textit{~Page D.,~Geppert U. and~Weber F.}
\Journal{\NPA}{777}{497}{2006}

\bibitem{Page:2004fy}
 \textit{~Page D.,~Lattimer J.~M.,~Prakash M. and~Steiner A.~W.}
\Journal{\APJS}{155}{623}{2004}

\bibitem{Blaschke:2004vq}
\textit{~Blaschke D.,~Grigorian H. and~Voskresensky D.~N.}
\Journal{\AA}{424}{979}{2004}


\bibitem{Voskresensky:2001fd}
\textit{~Voskresensky D.~N.}
\Journal{ Lect. Notes Phys.}{578}{467}{2001}

\bibitem{Yakovlev:2000jp}
\textit{~Yakovlev D.~G.,~Kaminker A.~D.,~Gnedin O.~Y. and ~Haensel
P.}
\Journal{\PREP}{354}{1}{2001}

\bibitem{Blaschke:2000dy}
 \textit{~Blaschke D.,~Grigorian H. and~Voskresensky D.~N.}
  \Journal{\AA}{368}{561}{2001}


\bibitem{Jaikumar:2001hq}
  \textit{~Jaikumar P. and~Prakash M.}
\Journal{\PLB}{516}{345}{2001}

\bibitem{Blaschke:1999qx}
 \textit{~Blaschke D.,~Kl\"ahn T. and~Voskresensky D.~N.}
  \Journal{\AJ}{533}{406}{2000}
  [arXiv:astro-ph/9908334].

\bibitem{Takatsuka:2004zq}
  \textit{~Takatsuka T. and~Tamagaki R.}
\Journal{\it  Prog.\ Theor.\ Phys.}{112}{37}{2004}

 \bibitem{Schwenk:2003bc}
 \textit{~Schwenk A. and~Friman B.}
\Journal{\PRL}{92}{082501}{2004}

\bibitem{Grigorian:2004jq}
 \textit{~Grigorian H.,~Blaschke D. and~Voskresensky D.}
\Journal{\PRC}{71}{045801}{2005}

\bibitem{Popov:2005xa}
\textit{~Popov S.,~Grigorian H. and~Blaschke D.}
\Journal{\PRC}{74}{025803}{2006}

\bibitem{Barrois:1977xd}
  \textit{~Barrois B.~C.}
\Journal{\NPB}{129}{390}{1977}.


\bibitem{Schafer:2000tw}
\textit{~Sch\"afer T.}
\Journal{\PRD}{62}{094007}{2000}

\bibitem{Alford:2002rz}
 \textit{~Alford M.~G.,~Bowers J.~A.,~Cheyne J.~M. and~Cowan
 G.~A.}
\Journal{\PRD}{67}{054018}{2003}


\bibitem{Aguilera:2005tg}
  \textit{~Aguilera D.~N.,~Blaschke D.,~Buballa M. and~Yudichev
  V.~L.}
\Journal{\PRD}{\bf 72}{034008}{2005}

\bibitem{Schmitt:2005wg}
  \textit{~Schmitt A.,~Shovkovy I.~A. and~Wang Q.}
\Journal{\PRD}{73}{034012}{2006}

\bibitem{Aguilera:2005uf}
 \textit{~Aguilera D.~N. and~Blaschke D.~B.}
  Phys.\ Part.\ Nucl.\ Lett.\ 2007. V.4. P.351.

\bibitem{Aguilera:2006cj}
\textit{~Aguilera D.~N.,~Blaschke D.,~Grigorian H. and ~Scoccola
N.~N.}
\Journal{\PRD}{\bf 74}{114005}{2006}


\bibitem{Blaschke:2004vr}
  D.~Blaschke, D.~N.~Voskresensky and H.~Grigorian,
{\it  Compact Stars. The Quest for New States of Dense Matter},
World Scientific, Singapore 2004. P.409.

  \bibitem{Grigorian:2005fd}
 \textit{~Grigorian H.}
\Journal{\PRC}{74}{025801}{2006}

\bibitem{Weber:1999qn}
 \textit{~Weber F.}
 {\it Pulsars as astrophysical laboratories for nuclear and particle physics},
 IOP Publishing, Bristol 1999.

\bibitem{Prakash:1996xs}
\textit{~Prakash M. et al.}
\Journal{\PREP}{280}{1}{1997}

\bibitem{Blaschke:2001uj}
 \textit{~Blaschke D.,~Glendenning N.~K. and~Sedrakian A.,  (Eds.):
  // Physics of neutron star interiors},
Springer, Lecture Notes in Physics, 2001. V.578.

\bibitem{Bombaci:2001uk}
 \textit{~Bombaci I.}
\Journal{Lect.\ Notes Phys.}{578}{253}{2001}

\bibitem{Li:1999wt}
  \textit{~Li X.~D.,~Bombaci I.,~Dey M.,~Dey J. and~van den Heuvel
  E.~P.~J.}
\Journal{\PRL}{83}{3776}{1999}


\bibitem{Ozel:2006bv}
 \textit{~\"Ozel F.} //
  Nature 2006. V.441. P.1115.


\bibitem{Glendenning:1997fy}
 \textit{~Glendenning N.~K.,~Pei S. and~Weber F.}
 \Journal{\PRL}{79}{1603}{1997}

\bibitem{Glendenning:2000zz}
\textit{~Glendenning N.~K. and~Weber F.}
 \Journal{\APJ}{559}{L119}{2001}

\bibitem{Poghosyan:2000mr}
 \textit{~Poghosyan G.~S.,~Grigorian H. and~Blaschke D.}
 \Journal{\APJ}{551}{L73}{2001}

\bibitem{Blaschke:2001th}
\textit{~Blaschke D.,~Bombaci I.,~Grigorian H. and
~PoghosyanG.~S.}
\Journal{\NA}{7}{107}{2002}

\bibitem{Blaschke:2006gd}
\textit{~Blaschke D. and~Grigorian H.}
 Prog. Part. Nucl. Phys. 2007.
doi:10.1016/j.ppnp.2006.12.035; [arXiv:astro-ph/0612092].


\bibitem{Kolomeitsev:2004ff}
 \textit{~Kolomeitsev E.~E. and~Voskresensky D.~N.}
\Journal{\NPA}{759}{373}{2005}

\end{thebibliography}
\end{document}